\documentclass[draft,
	twocolumn,
preprintnumbers,
amsmath,
	amssymb,
	aps,
pre,
floatfix,
	fleqn,
]{revtex4-1}

\usepackage{dynlearn}
\usepackage{subfig}

\graphicspath{{./imp/}}
\usepackage{physics} \usepackage{multirow}
\usepackage{cleveref}
\usepackage{nccmath}

\newcommand{\defn}{\stackrel{.}{=}} \newcommand{\kb}{k_\textnormal{B}} \newcommand{\ex}{\mathrm{e}} \newcommand{\param}{\alpha} \newcommand{\xx}{x} \newcommand{\xsp}{\mathcal{X}} \newcommand{\ST}{X} \newcommand{\mus}{\mu} \newcommand{\mud}{\bm{\mus}} \newcommand{\ssde}{\pi} \newcommand{\ssdb}{\bm{\ssde}} \newcommand{\Gmat}{\mathbf{G}} \newcommand{\Tmat}{\mathbf{T}} \newcommand{\Wex}{W_\textnormal{ex}} \newcommand{\Qex}{Q_\textnormal{ex}} \newcommand{\Qhk}{Q_\textnormal{hk}} \newcommand{\gWex}{\mathcal{W}_\textnormal{ex}} \newcommand{\gQex}{\mathcal{Q}_\textnormal{ex}} \newcommand{\gQhk}{\mathcal{Q}_\textnormal{hk}}  \newcommand{\Feq}{F_\param^{\textnormal{eq}}}     \newcommand{\GNa}{\Gmat_\param^\textnormal{Na$^+$}}
\newcommand{\GK}{\Gmat_\param^\textnormal{K$^+$\phantom{a}}}
\newcommand{\Fnss}{\mathcal{F}^{\textnormal{nss}}}

\newcommand{\pr}[1]{\mathrm{Pr}\left(#1\right)} \newcommand{\cpr}[2]{\mathrm{Pr}\left(#1 \mid #2\right)}   \renewcommand{\DKL}[2]{D_\textnormal{KL}\bqty{#1 \, \| \, #2}}  
\newcommand{\ssda}[1]{ \ssde_{\param} \pqty{#1} }

\newcommand{\aGmat}[1]{\mathbf{\Gmat}_{\param_{#1}}}  \newcommand{\Taij}[1]{\bqty{\Tmat_{\param}^{\Delta t}}_{#1}}

\newcommand{\FP}[2]{\mathcal{P}_{#2}\left( #1 \right)}

\newcommand{\RP}[2]{\mathcal{R}_{#2}\left( #1 \right)}

\newcommand{\muF}{\bm{\mu}_\textnormal{F}}
\newcommand{\muR}{\bm{\mu}_\textnormal{R}}

\begin{document}

\def\ourTitle{Homeostatic and adaptive energetics:\\
Nonequilibrium fluctuations beyond detailed balance in
voltage-gated ion channels
}

\def\ourAbstract{Stochastic thermodynamics has largely succeeded in characterizing both
equilibrium and far-from-equilibrium phenomena. Yet many opportunities remain
for application to mesoscopic complex systems---especially biological
ones---whose effective dynamics often violate detailed balance and whose
microscopic degrees of freedom are often unknown or intractable. After reviewing
\emph{excess} and \emph{housekeeping} energetics---the adaptive and homeostatic
components of a system's dissipation---we extend stochastic thermodynamics with
a trajectory class fluctuation theorem for nonequilibrium steady-state,
nondetailed-balanced complex systems. We then take up the neurobiological
examples of voltage-gated sodium and potassium ion channels to apply and
illustrate the theory, elucidating their nonequilibrium behavior under a
biophysically plausible action potential drive. These results uncover challenges
for future experiments and highlight the progress possible understanding the
thermodynamics of complex systems---\emph{without} exhaustive knowledge of every
underlying degree of freedom.
}

\def\ourKeywords{rate equations, stochastic process, hidden Markov model, information processing, logical circuits, ion channel, entropy production, reversibility
}

\hypersetup{
  pdfauthor={Mikhael Semaan},
  pdftitle={\ourTitle},
  pdfsubject={\ourAbstract},
  pdfkeywords={\ourKeywords},
  pdfproducer={},
  pdfcreator={}
}

\author{Mikhael T. Semaan}
\email{msemaan@ucdavis.edu}
\affiliation{Complexity Sciences Center and Department of Physics and Astronomy, University of California, Davis, One Shields Avenue, Davis, CA 95616}

\author{James P. Crutchfield}
\email{chaos@ucdavis.edu}
\affiliation{Complexity Sciences Center and Department of Physics and Astronomy, University of California, Davis, One Shields Avenue, Davis, CA 95616}

\bibliographystyle{unsrt_abbrv_custom}

\title{\ourTitle}

\begin{abstract}
  \ourAbstract
\end{abstract}


\preprint{\arxiv{2202.13038}}

\title{\ourTitle}
\date{\today}
\maketitle

\setstretch{1.0}

\section{Introduction}
\label{sec:intro}

Nonequilbrium phenomena pervade nature: In their many forms, energy gradients
send hurricanes and wildfires to ravage, volcanoes to form and erupt, life to
emerge. Mesoscopic complex systems---a planetary climate, forest ecosystems, the
human body---consist of microscopic degrees of freedom that are inaccessible,
intractable, or simply unknown. In point of fact, the human body's biochemistry
relies essentially on out-of-equilibrium dynamics to function, adapt, and
maintain homeostasis; its myriad degrees of freedom are only ever partially
accessible. Similarly, mesoscopic and complex systems provide fertile grounds
for honing and applying tools to analyze real-world nonequilibrium processes.

Describing energetic fluxes in complex systems---developing a suitable
mesoscopic nonequilibrium thermodynamics---remains an ongoing challenge:
mathematics and physics difficulties continue to hinder deeper understanding of
how these systems operate and function. The following leverages and extends
tools from stochastic thermodynamics and information theory to address these
challenges. To demonstrate the techniques, it takes up two suitably complex,
mesoscopic neurobiological systems: voltage-gated ion channels.

\subsection{Nonequilibrium steady states}

A system is typically called nonequilibrium in two distinct senses. The first,
and most common, refers to nonequilibrium \emph{processes}---say, induced by
rapid environmental driving---wherein a system evolves through a series of
transient configurations. When the environmental drive remains fixed, such a
system remains out of equilibrium as it \emph{relaxes} to some stationary
distribution over its states, determined by the environmental parameters. If
that stationary distribution corresponds to a thermodynamic equilibrium, we say
the system possesses an \emph{equilibrium steady state} (ESS), irrespective of
its (perhaps highly nonequilibrium) driven, transient dynamics.

The second sense refers not to the transient behavior but to the nature of the
stationary distributions: a \emph{nonequilbrium steady-state} (NESS) system is
one whose steady states are themselves out of thermodynamic equilibrium. This is
simply achieved by contact with two heat baths at different
temperatures. Rayleigh-B\'enard convection \cite{Cros93a} exemplifies this
phenomenon: the temperature gradient between the top and bottom boundaries
ensures a constant flux of energy through the fluid, from the hotter to the
cooler, even when the gradient remains fixed indefinitely. In this case it is
not enough to identify the energetic fluxes due to the system's transient
dynamics; we must also identify the energy required to maintain steady-state
conditions in the first place.

NESSs appear even without multiple heat baths. For example, by optically
dragging a bead through viscous fluid
\cite{trepagnierExperimentalTestHatano2004}---an experimental realization of
nonconservative force-driven Langevin dynamics---by coarse-graining microstates
\cite{espositoStochasticThermodynamicsCoarse2012}; or by contact with reservoirs
of distinct electrochemical potentials---the case in virtually all common
electrical circuits via Joule heating \cite{Gao21a}. They emerge as well in the
voltage-gated ion channels we consider.

A first attempt to give NESS systems a full thermodynamic framing defined the
\emph{housekeeping} heat $\Qhk$ as the portion of the total heat $Q$ that
maintains NESS conditions \cite{oonoSteadyStateThermodynamics1998}. (In this,
the total heat is that energy exchanged between a system and its thermal
environment, often idealized as a fixed-temperature bath.) What remains is the
energy exchanged owing to the system's relaxation to steady state, termed the
\emph{excess} heat $\Qex$:
\begin{align}
  Q = \Qex + \Qhk
  \textnormal{.}
\label{eq:Q_breakdown}
\end{align}

Contrast this with an equilibrium system's steady states, which by definition
exchange no net energy with the thermal environment. In this setting, $\Qhk = 0$
and so \emph{all} dissipated heat is excess: $\Qex \to Q$. In other words,
$\Qex$ in the NESS setting carries the same meaning as \emph{total} heat $Q$ in
the ESS setting, and vice versa.

\subsection{Approach}

Equilibrium thermodynamics and equilibrium statistical mechanics prove
insufficient to analyze nonequilibrium processes \footnote{With important
exceptions, particularly in the first sense, under small perturbations from
equilibrium \cite{Onsa31a,Kond08a}.}. That said, recent advances in stochastic
thermodynamics now successfully describe fluctuations in a variety of
far-from-equilibrium systems. This has been done both in the first sense
(relaxation to ESSs)
\cite{jarzynskiEquilibriumFreeenergyDifferences1997,jarzynskiNonequilibriumEqualityFree1997,crooksEntropyProductionFluctuation1999,crooksNonequilibriumMeasurementsFree1998}
and in the second (NESSs)
\cite{hatanoSteadyStateThermodynamicsLangevin2001,speckIntegralFluctuationTheorem2005,riechersFluctuationsWhenDriving2017}.
Ref.~\cite{seifertStochasticThermodynamicsPrinciples2018} gives a recent review.

The following applies and extends these advances to analyze two complex
neurobiological systems: voltage-gated sodium and potassium ion channels
\cite{dayanTheoreticalNeuroscienceComputational2001}---biophysical systems that
originally motivated introducing master equations for NESSs
\cite{schnakenbergNetworkTheoryMicroscopic1976}. This elucidates, for the first
time, their nonequilibrium behavior under the realistic, dynamic environmental
drive of an action potential spike. In doing so, a toolkit emerges whose
validity extends to a host of other mesoscopic complex systems---even those for
which a purely energetic interpretation is impossible or problematic---provided
a relatively small set of constraints on their effective dynamics.

Our development unfolds as follows. First, Sec.~\ref{sec:prelim} lays out the
relevant notation for our model classes and introduces appropriate \emph{excess}
thermodynamic functionals for describing them, ending with
Sec.~\ref{subsec:db_housekeeping} which elaborates on the relationship between
housekeeping heat, (ir)reversibility, and detailed balance. Sec.~\ref{sec:FTFE}
reviews \emph{fluctuation theorems}, which bind nonequilibrium thermodynamic
fluctuations to steady-state quantities. It closes in
Sec.~\ref{subsec:TCFT_theory} with our primary theoretical result: the first
full trajectory class fluctuation theorem valid for NESS systems.

Moving to applications, Sec.~\ref{sec:ion_intro} introduces our example
neurobiological systems: voltage-gated sodium and potassium ion channels
embedded in neural membranes. Sec.~\ref{sec:results} then applies the techniques
developed in the preceding theory to the channels, illustrating and comparing
their responses under realistic action potential spikes.

These results serve three roles. First, they show how the trajectory class
fluctuation theorem evades the divergences implied by real-world systems with
one-way only transitions. Second, they quantitatively demonstrate how failing
to account for housekeeping dissipation violates related fluctuation theorems,
suggesting an important direction for experimental effort. Finally, despite
marked differences between the ion channels' steady-states, the results show
how to directly compare the channels' \emph{excess} energetics. This both
circumvents implied housekeeping divergences and allows for meaningful
comparisons between their \emph{adaptive} responses to the same environmental
stimulus.

\section{Preliminaries}
\label{sec:prelim}

\begin{figure}[ht]
\centering
\includegraphics[width=\columnwidth]{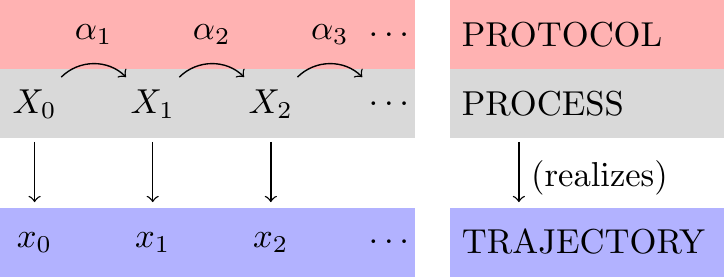}
\caption{Interaction between the stochastic process $X_{0:N}$, protocol
	$\param_{0:N}$, and realized (observed) trajectories $x_{0:N}$.
	}
\label{fig:process_visual}
\end{figure}

The central object here is the finite-length \emph{controlled stochastic
process} $X_{0:N} \defn X_{0} X_{1} X_{2} \dots X_N$, where $X_i \in
\mathcal{X}$ is the random variable corresponding to the state of a system under
study (SUS) at times $\{ t_i \in \mathbb{R}: i = 0,\ldots,N\}$. We call a
specific realization $x_{0:N}$ a \emph{trajectory}. The process' dynamics are
not stationary; rather, they are driven by a \emph{protocol} $\param_{0:N}$.
Fig.~\ref{fig:process_visual} illustrates the scheme.

We place the following constraints on the SUS.
\begin{enumerate}[leftmargin=*]
      \setlength{\topsep}{-2pt}
      \setlength{\itemsep}{-2pt}
      \setlength{\parsep}{-2pt}
\item Each system parameter $\param_i$ leads to a stationary and ergodic
	stochastic process, realized by holding the protocol fixed indefinitely at
	$\param_i$. This implies a unique would-be steady-state distribution
	$\bm{\pi}_{\param_i}$ associated to each $\param_i$.
\item The state and protocol spaces are of \emph{even parity}, in the sense that
	we do not negate their values under time reversal, defined precisely later.
	Sec.~\ref{subsec:db_housekeeping} discusses removing this assumption.
\end{enumerate}

We emphasize these are \emph{all} that is required for the main theoretical
result and for meaningful definitions of the excess and housekeeping
functionals. Importantly, we do not require dynamics of any particular form or
possessing any particular structure---Markovian, Langevin, detailed-balanced,
Hamiltonian, master equation, coupled to ideal baths, and so on---beyond that
specified by the two conditions above. We do not require states to be
\emph{microscopic}; they can correspond to arbitrary or unknown coarse
grainings. With this in mind, even the discrete succession of events is
flexible. In particular, from any continuous-time dynamic we may generate a
corresponding discrete-time one for appropriately small time steps.

While one cannot, in this most general setting, determine \emph{energetics},
the fluctuation theorems introduced hold independently and exactly---and at any
level of system description. We state the fluctuation theorems in this setting
for two primary reasons: first, for clarity of derivation; second, with an eye
toward future applications beyond thermodynamic systems to generally
nonstationary stochastic processes.

\subsection{The thermodynamic system}
\label{subsec:thermo_cartoon_system}

\begin{figure}[ht]
\centering
\includegraphics[width=\columnwidth]{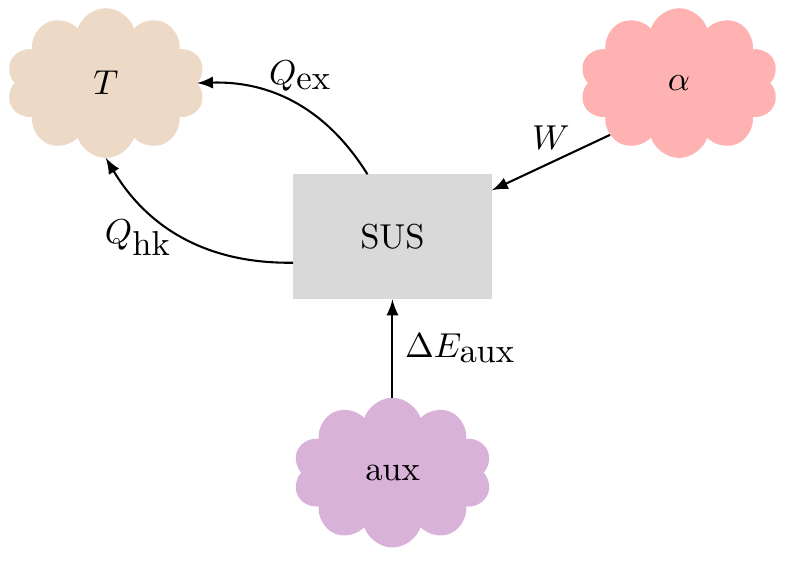}
\caption{A \emph{thermal system} and its interactions with various baths. While
	heat and work reservoirs (labeled with temperature $T$ and parameter $\param$,
	respectively) are ideal, by design we assume nothing about the auxiliary or
	``aux'' reservoir, and so label its energetic contribution by $\Delta
	E_\textnormal{aux}$ to avoid confusion with well-defined terms like heat and
	work. The labels and arrow directions indicate energetic fluxes to-from the
	system. Notably, we allow for nonequilibrium steady states and functionally
	split the total heat $Q$ into the \emph{excess} heat $\Qex$---corresponding to
	\emph{adaptive} dissipation---and \emph{housekeeping} heat $\Qhk$---referring
	to \emph{homeostatic} dissipation.
	}
\label{fig:thermo_visual}
\end{figure}

That said, generality can hinder ease of application. To this end, when
presenting the theoretical tools we frequently return to the relevant example
``thermal system'' of Fig.~\ref{fig:thermo_visual}. This is a SUS coupled to an
ideal heat bath at inverse temperature $\beta = 1/\kb T$, an ideal work
reservoir parameterized by $\param$, and an \emph{auxiliary reservoir}
representing the otherwise unaccounted-for degrees of freedom. Furthermore, we
assign to each SUS state $x$ an energy $E_\param \pqty{x}$. Finally, while the
example system does not assume (order-$1$) Markov dynamics, it \emph{does}
assume no dynamical dependency on times before $t_0$. That is, the system's
initial preparation is sufficient to determine the stochastic dynamics during
the protocol.

These additional restrictions allow identifying the \emph{energies} associated
to each dynamical functional---introduced shortly. While these constraints are
minimal, they still allow the SUS to be a coarse-grained representation. In
general, this implies that the energetic fluxes are bounds rather than strict
equalities \cite{espositoStochasticThermodynamicsCoarse2012}.

Of paramount importance---and missing from most idealized thermal schemes---is
the presence of the auxiliary bath. The heat and work reservoirs are each
proxies for distinct \emph{kinds} of coarse-grained degrees of freedom with
distinct internal structures: the heat reservoir is an infinitely large source
of purely thermal energy; the work reservoir is an entropyless source of energy,
whose role is to set the SUS's energetic landscape via parameters $\param$.

In contrast, there are no restrictions on the auxiliary reservoir's structure.
It is unnecessary when describing the SUS's effective state (or energy) at any
particular time. Partly, the auxiliary reservoir stands in for coarse-graining
out unknown degrees of freedom by unknown schemes. In the case of
Rayleigh-B\'enard convection, the  auxiliary reservoir is a heat bath at a
different temperature $T_2$. To take another example, in an information ratchet
scheme, the auxiliary reservoir may represent the information tape interacting
with the ratchet system
\cite{boydIdentifyingFunctionalThermodynamics2016,boydThermodynamicsModularityStructural2018,jurgensFunctionalThermodynamicsMaxwellian2020}.

Yet the relevance of the auxiliary bath goes beyond this:  specifically,
Sec.~\ref{subsec:db_housekeeping} shows that it is a \emph{necessary} source for
maintaining NESS conditions in this idealized picture.

Altogether, Fig.~\ref{fig:thermo_visual} captures a large class of mesoscopic
physical, chemical, biological, and engineered systems that exhibit
nonequilibrium steady states, but that have additional structure and are in
contact with at least one thermal environment. We note an important distinction:
the ``heats'' to which we refer in the thermal system context are always
associated to the heat bath defined in Fig.~\ref{fig:thermo_visual}, and
``works'' associated to $\param$. We avoid calling any fluxes between the
auxiliary bath and the system heat or work, since we place no \emph{a priori}
restriction on its structure.

Subsequent sections develop tools for calculating the associated heats and works
and for bounding their nonequilibrium fluctuations. Practically, this suggests
experimental calorimetry and introduces a valuable way to \emph{calibrate}
effective models---such as, for example, those of the ion channels we take up
later.

\subsection{Excess energetics}
\label{subsec:core_definitions}

For an ESS system in contact with a single heat bath, the familiar First Law
defines work $W$ and heat $Q$ as distinct contributions to its energy change
over the course of a protocol \cite{jarzynskiNonequilibriumEqualityFree1997}:
\begin{alignat}{2}
  \Delta E &= \underbrace{ \int \partial_\param E_\param
  \pqty{\xx} \dd{\param} } &&+ \underbrace{ \int \partial_\xx E_\param
  \pqty{\xx} \dd{\xx} } \notag \\
  &\defn \qquad \; \; W && \qquad \; \;  - Q
\textnormal{.}
\label{eq:1L_energy} 
\end{alignat}

That is, $W$ denotes a difference in system energy owing to a change in
protocol---a change in the overall energy \emph{landscape}---and $Q$ denotes the
difference owing to a change in system state---a dissipative signature of its
\emph{adaptation} to environmental conditions.

In contrast, the general setting may not provide a meaningful notion of energy.
Worse, even in the restricted case of Fig.~\ref{fig:thermo_visual}, we can no
longer define total heat for a NESS system by Eq.~\eqref{eq:1L_energy}: it leads
to contradiction.

To see this, consider a fixed protocol at $\param$ and a system poised already
in the distribution $\bm{\pi}_\param$. By definition then $W = 0$ and so
$\expval{W} = 0$, where $\expval{\cdot}$ denotes a weighted average over all
possible paths. We also have $\expval{\Delta E} = \Delta
\braket{\pi_\param}{E_\param} = 0$, where:
\begin{align}
  \braket{\pi_\param}{E_\param} &\defn
    \int_{x \in \mathcal{X}} \pi_\param\pqty{x} E_\param\pqty{x} 
    \textnormal{.}
\label{eq:state_avg_defn}
\end{align}
Yet we cannot have $\expval{Q} = 0$, for by Eq.~\eqref{eq:Q_breakdown}:
\begin{align}
  \expval{Q} & = \expval{\Qhk} \nonumber \\
  & \neq 0
  ~,
\label{eq:avg_Qhk_neq_0}  
\end{align}
since this is a NESS system. In other words, the observed housekeeping flux,
leaving no signature in the state energies, must come from somewhere
\emph{outside} of the system coupled to single ideal heat and work reservoirs.
This is precisely what the auxiliary bath provides: in this case $\Delta
E_\textnormal{aux} = \Qhk$.

However, there is an alternative to energy. Since to every parameter $\param$
is an associated steady-state distribution $\bm{\pi}_\param$, we can define the
\emph{steady-state surprisal}:
\begin{align}
\phi_\param \pqty{\xx} \defn - \ln \ssde_\param \pqty{\xx}
  ~;
\label{eq:ss_surprisal}
\end{align}
so called since it is the Shannon self-information
\cite{coverElementsInformationTheory2006} of observing state $x$ under the
distribution $\ssdb_\param$.

Taking the surprisal's state average under this distribution yields its Shannon
entropy (or, for continuous-state spaces, its differential entropy):
\begin{align}
\braket{\pi_\param}{\phi_\param} &\defn
\int_{x \in \mathcal{X}} \pi_\param\pqty{x} \phi_\param\pqty{x} 
  \notag \\
  &= H\bqty{\ssdb_\param}
  ~.
\label{eq:shannon_ssd}
\end{align}

To see how the surprisal relates to energy, consider the canonical ensemble of
statistical mechanics---the ESS version of Fig.~\ref{fig:thermo_visual}, with
$\Qhk = 0$ and so no auxiliary bath---where $\ssdb_\param$ is the Boltzmann
distribution \cite{sethnaEntropyOrderParameters2021}. Then:
\begin{align}
- \ln \ssde_\param \pqty{\xx} &= \beta \pqty{E_\param \pqty{\xx} - \Feq} 
  \nonumber \\
  & = \phi_\param \pqty{\xx}
  ~,
\label{eq:Boltzmann}
\end{align}
where $\Feq$ is the equilibrium free energy (the familiar logarithm of the
canonical partition function).

Eq.~\eqref{eq:Boltzmann} motivates yet another moniker for $\phi_\param
\pqty{x}$: the \emph{nonequilibrium potential}. In this sense, steady-state
surprisal is analogous to a generalized energy. However, it remains a
meaningful characterization of a system's steady-state distribution---via its
information-theoretic interpretation---even when energy is not meaningful.

Leveraging this, an analogous First Law for $\phi_\param\pqty{x}$ defines the
\emph{excess} heat and work:
\begin{alignat}{2}
  \Delta \phi &= \underbrace{ \int \partial_\param \phi_\param
  \pqty{x} \dd{\param} } &&+ \underbrace{ \int \partial_x
  \phi_\param \pqty{x} \dd{x} } \notag \\
  &\defn \quad \quad \; \; \gWex && \qquad \; \, - \gQex
  \textnormal{.}
\label{eq:1L_phi}
\end{alignat}
As with their nonexcess counterparts, these quantities characterize distinct
\emph{dynamical} contributions to a change in steady-state surprisal: $\gWex$
capturing that due to a changing protocol, which sets the steady-state
probability landscape; $\gQex$ monitoring a system's \emph{adaptation} to its
environment.

For Fig.~\ref{fig:thermo_visual}'s thermal system, these conveniently convert
to energies: $\gWex \to \beta \Wex$ and $\gQex \to \beta \Qex$. And, they agree
with other standard formulations of excess thermodynamic functionals
\cite{hatanoSteadyStateThermodynamicsLangevin2001,riechersFluctuationsWhenDriving2017}.
Using Eq.~\eqref{eq:Boltzmann} and taking the ESS limit of
Boltzmann-distributed steady-states yields: (i) $\gQex \to \beta Q$---with
equilibrium steady states, all dissipated heat is excess---and (ii) $\gWex \to
\beta \pqty{W - \Delta F}$, leading to its classification as an \emph{excess
environmental entropy production} \cite{riechersFluctuationsWhenDriving2017}.

We stress, though, that the excess work and heat---and the steady-state
surprisal---retain dynamical meaning independent of Boltzmann or even energetic
assumptions. In this way, Eq.~\eqref{eq:Boltzmann} is a guidepost for
thermodynamic interpretation. It is not, however, a strict equivalence. In
point of fact, as we will see, sodium channels (as with other NESS systems)
lack a well-defined steady-state free energy
\cite{riechersFluctuationsWhenDriving2017}. Nevertheless, Eq.~\eqref{eq:1L_phi}
describes---tractably---two functionally distinct aspects of their response to
dynamic environments.

\subsection{Detailed balance and housekeeping}
\label{subsec:db_housekeeping}

The housekeeping heat remains. Recall that it corresponds to energy dissipated
to maintain NESSs, as in Eq.~\eqref{eq:Q_breakdown}. Phenomenologically,
Eq.~\eqref{eq:Q_breakdown} provided a satisfactory answer. However, our excess
heat definition only required would-be steady-state distributions exist. The
definition of total heat, in contrast, depended explicitly on well-defined state
energies. This difference led to problems with NESSs.

The upshot is that a more general definition of housekeeping heat is called
for. In particular, it should depend only on the stochastic dynamics and, when
added to excess heat, it should give a reasonable generalization of
\emph{total} heat. Naturally, we also require interpretability and that it
reduces to the corresponding well-understood thermodynamic terms in the
appropriate limits.

To these ends, but in a slightly more general form than previously reported, we
define \emph{housekeeping heat} to explicitly allow non-Markovian dynamics:
\begin{align}
  \gQhk \defn \ln \,&\frac{
    \cpr{X_{1:N} = x_{1:N\phantom{-1}}}{X_0 = x_{0\phantom{N}}; \param_{0:N}}
  }{
    \cpr{X_{1:N} = x_{N-1:0}}{X_0 = x_{N\phantom{0}}; \param_{N:0}}
  } \notag \\
  &+ \ln \prod_{i = 0}^{N-1} \frac{
    \pi_{\param_{i+1}}\pqty{_{\phantom{+}} x_{i\phantom{1}}}
  }{
    \pi_{\param_{i+1}}\pqty{x_{i+1}}
  }
  \textnormal{.}
\label{eq:Qhk_general}
\end{align}
Observe that the first term is a log-ratio of conditioned path probabilities.
The denominator is the numerator's time reversal: the probability
of obtaining the reversed path $x_{N:0}$ conditioned on starting in the state
$x_N$ and subject to the reversed protocol $\param_{N:0}$. The second term is
exactly $-\gQex$ by the discrete form of Eq.~\eqref{eq:1L_phi}. And so, by
identifying $\mathcal{Q} \defn \gQex + \gQhk$, housekeeping heat is a component
of the generalized \emph{total heat} $\mathcal{Q}$.

In the (single heat bath) thermal example, one recovers units of energy as
$\mathcal{Q} \to \beta Q$ and $\gQhk \to \beta \Qhk$. And, the resulting total
heat is consistent with formulations based on \emph{microscopic reversibility}
\cite{crooksEntropyProductionFluctuation1999}. Equivalently, we could have
started with this microscopic reversibility condition for even state spaces and
arrived at the appropriate housekeeping heat.

With this in mind, consider \emph{relaxing} the even state space assumption.
Doing so and keeping the appropriate microscopic reversibility condition allows
for an analogous splitting of housekeeping heat---modified so that the
denominator's terms are negated where required---and excess heat, consistent
with previous considerations of odd-parity NESS systems
\cite{spinneyNonequilibriumThermodynamicsStochastic2012,yeoHousekeepingEntropyContinuous2016}.
While we note that an analogue to our Eq.~\eqref{eq:NESS_TCFT} holds, we do not
treat this further here.

Now, consider a Markov dynamic of order $1$. That is, conditioning on the
previous time step fully characterizes the probability distribution over
futures. Then, the first term reduces to the logarithm of a product of one-step
conditional probabilities. And, $\gQhk$ tracks the degree of
\emph{detailed-balance violation} over the trajectory. This is in agreement
with existing definitions
\cite{hatanoSteadyStateThermodynamicsLangevin2001,harrisFluctuationTheoremsStochastic2007,
riechersFluctuationsWhenDriving2017}. Concretely, detailed-balanced dynamics
imply $\gQhk = 0$ for every trajectory. If any trajectory yields $\gQhk \neq
0$, the dynamic is necessarily nondetailed-balanced.

Finally, recall that by definition $\gQhk = 0$ for an ESS system. Taken
together with assuming an even state space---ensuring correct ``reverse''
probabilities---the Markov condition says, succinctly:
\begin{align*}
  &\textnormal{\emph{nondetailed-balanced dynamics}} \\
  &\qquad \Updownarrow \\
  &\textnormal{\emph{nonequilibrium steady states}.}
\end{align*}
Recall that the Markov condition is appropriate for many microscopically-modeled thermal systems such as overdamped Langevin dynamics, as well as for a host of biological systems like the ion channels we consider later.

Nonzero housekeeping heat actually \emph{necessitates} including an auxiliary
reservoir for a complete picture. Recall Fig.~\ref{fig:thermo_visual}. This
follows since a NESS system, even fully relaxed to its stationary distribution,
constantly dissipates housekeeping heat to the thermal reservoir. (And does so
at an average rate of $\dd\expval{\Qhk} / \dd t$.) Yet, with the protocol
parameter fixed, no work (or excess work) is done: $W = 0$ by
Eq.~\eqref{eq:1L_energy}. The system's average energy does not change, though,
since the parameter and individual state uniquely set its energies:
$\dd\expval{E} / \dd t = 0$.

The conclusion is that energy flux through the system, observed in the
housekeeping dissipation to the thermal reservoir, must come from
\emph{somewhere} not otherwise described by the ideal constructs. In other
words, in the thermodynamically-interpretable setting, nondetailed-balanced
dynamics are signatures of unaccounted-for degrees of freedom. In this way, the
constructions in Eqs.~\eqref{eq:1L_phi} and \eqref{eq:Qhk_general} provide the
tools to isolate this \emph{homeostatic} part of a system's energetic fluxes,
so called for its role maintaining homeostatic (steady-state) conditions.

We close by calling out a feature on direct display in
Eq.~\eqref{eq:Qhk_general}. While placing minimal restrictions on the
\emph{dynamics}, problems arise when any path is strictly irreversible, in the
sense that a nonzero-probability forward trajectory is associated a
zero-probability reverse. Then, $\gQhk$ diverges. And this seemingly forbids
dynamics in finite state spaces with one-way-only transitions.

In the thermodynamic interpretation, such a transition costs infinite
dissipation. And, with this realization, usually a model's mesoscopic nature
comes to bear. Indeed, Ref.~\cite{riechersFluctuationsWhenDriving2017} in its
related sodium channel analysis remarks that ``more careful experimental effort
should be done to bound the actual housekeeping entropy production in these ion
channels''. The following section demonstrates that a new
\emph{trajectory class fluctuation theorem} provides a tool for analyzing such
experiments and circumvents the divergence while still placing strong bounds on fluctuations.

\section{Fluctuations and Free Energy}
\label{sec:FTFE}

So far, we defined the generalized quantities $\gWex$, $\gQex$, and $\gQhk$ and
elucidated their meanings outside the ESS regime. As with their ESS
counterparts, though, they depend on the specific path a system takes through
its state space under a particular protocol. A suite of statistical tools
called \emph{fluctuation theorems} (FTs) tie such nonequilibrium behaviors to
equilibrium (or steady-state,  more generally) quantities. They come in three
primary flavors: (i) \emph{integral} FTs (IFTs) concern weighted averages over
all possible trajectories, (ii) \emph{detailed} FTs (DFTs) fix the relationship
between a specific path and its associated reversal, and (iii) \emph{trajectory
class} FTs (TCFTs) interpolate between the two \cite{Wims19a}.

The remainder of this section compares and contrasts these, discusses their
relation to free energy, and concludes with a TCFT for NESS
systems. This sets the stage for analyzing the two ion channels' thermodynamic
responses---expressed in terms of excess work, excess heat, and housekeeping
heat---to complex environmental signals.

\subsection{Fluctuation theorems}
\label{sec:FTs}

Integral and detailed FTs each exhibit complementary tradeoffs---tradeoffs
discussed below as we introduce the theorems. TCFTs, meanwhile, combine the
strengths of both and so are adaptable to a variety of systems and experimental
conditions. Here we present a general TCFT valid for NESS systems. It
simultaneously extends the previously known ESS FT and reveals experimental
difficulties unique to NESS systems, ultimately suggesting a need for new
experimental tools.

Jarzynski's equality \cite{jarzynskiEquilibriumFreeenergyDifferences1997,
jarzynskiNonequilibriumEqualityFree1997}, an IFT and the progenitor of the FTs
we consider, links equilibrium free energies to the averaged exponential work
distribution. It applies specifically to ESS systems that begin in their
equilibrium distribution and are connected to a single heat bath. Under these
conditions:
\begin{align}
  \expval{\ex^{-\beta W}} = \ex^{-\beta \Delta F^\textnormal{eq}}
  ~,
\label{eq:jarzynski}
\end{align}
where the angle brackets again refer to a weighted average over all possible
trajectories. That is, Jarzynski's equality ties an arbitrarily nonequilibrium
quantity---the averaged exponential work $\expval{\ex^{-\beta W}}$---to the
equilibrium free energy difference $\Delta F^\textnormal{eq}$---a state
function. Practically, this enables free energy estimation from nonequilibrium
work measurements \cite{cetinerRecoveryEquilibriumFree2020}.

It comes with disadvantages, however. In particular, extremely rare paths often
dominate the exponential work distribution
\cite{jarzynskiRareEventsConvergence2006}, leading to poor statistical accuracy
when estimating with finitely many experimental realizations. Nonetheless,
Jarzynski's equality has been confirmed for a wide variety of systems
\cite{hummerFreeEnergyReconstruction2001,liphardtEquilibriumInformationNonequilibrium2002,jarzynskiWorkFluctuationTheorems2006}.
In addition, while Eq.~\eqref{eq:jarzynski} only applies to ESS systems, a
variety of generalizations have been derived and tested for NESS systems
\cite{hatanoSteadyStateThermodynamicsLangevin2001,trepagnierExperimentalTestHatano2004,
speckIntegralFluctuationTheorem2005,seifertStochasticThermodynamicsFluctuation2012}.

In contrast to Jarzynski's IFT, the \emph{detailed} FTs (DFTs), express a
symmetry relation between a particular trajectory-protocol pair and its
appropriate time reversal. Perhaps the most well-known of these is due to
Crooks \cite{crooksNonequilibriumMeasurementsFree1998,
crooksEntropyProductionFluctuation1999}, which is complementary to Jarzynski's
IFT in several ways. For one, it makes the same assumptions: an ESS system
connected to a bath, beginning in equilibrium and driven away from it. For
another, Jarzynski's IFT results directly from trajectory-averaging both sides
of Crooks' DFT. Before presenting the DFT, though, we pause to precisely define
and set notation for what we mean by an ``appropriate reversal''.

Consider a system that begins in state distribution $\muF$, is driven by the
protocol $\param_{1:N}$, and realizes a trajectory in the measurable subset $C
\subseteq \xsp^{N+1}$. We call $C$ a \emph{trajectory class}. Then, we define
the \emph{forward process probability} as:
\begin{align}
\FP{C}{\muF} \defn \cpr{C}{ X_0 \sim \muF; \param_{1:N} } 
  ~.
\label{eq:FP_defn}
\end{align}
(Here, $\sim$ means ``is distributed as''.)
Now, consider the same system beginning in the distribution $\muR$ and driven
by the \emph{reverse protocol} $\widetilde{\param}_{N:1}$, where the tilde
indicates negation of time-odd variables (such as magnetic field). In turn,
we define the \emph{reverse process probability}:
\begin{align}
  \RP{C}{\muR} \defn 
  \cpr{C}{ X_0 \sim \muR; \widetilde{\param}_{N:1} } 
  \textnormal{.} 
\label{eq:RP_defn}
\end{align}
For finite state spaces, Eqs.~\eqref{eq:FP_defn} and \eqref{eq:RP_defn} define
distinct probability measures on the trajectory space. In a continuous state
space, we use the same notation to indicate probability \emph{densities}.

Let $\bm{\pi}_\textnormal{F} \defn \bm{\pi}_{\param_0}$ and $\bm{\pi}_\textnormal{R} \defn \bm{\pi}_{\widetilde{\param}_{N}}$. In these terms, Crooks' DFT reads:
\begin{align}
  \frac{
    \FP{x_{0:N}}{\bm{\pi}_\textnormal{F}}
  }{
    \RP{\widetilde{x}_{N:0}}{\bm{\pi}_\textnormal{R}}
  }
  = \ex^{\beta \pqty{W - \Delta F^\textnormal{eq}}}
\textnormal{.} 
\label{eq:crooks_DFT}
\end{align}
As with Jarzynski's IFT, the Crooks DFT has withstood experimental test
\cite{collinVerificationCrooksFluctuation2005} and seen use in empirically
estimating free energy differences \cite{cetinerRecoveryEquilibriumFree2020}.
Also, paralleling Jarzynski's IFT, Crooks' DFT has been generalized to a
variety of NESS systems
\cite{lahiriFluctuationTheoremsExcess2014,mandalEntropyProductionFluctuation2017,riechersFluctuationsWhenDriving2017}.

We highlight Ref.~\cite{riechersFluctuationsWhenDriving2017}'s generalization
of these. We recall, in particular, its Eq.~(25), since it is the DFT upon
which we base our TCFT.

Here and in the remainder, we assume even state and protocol
spaces (keeping in mind Sec.~\ref{subsec:db_housekeeping}'s notes on relaxing
this assumption), so there is never negation under time reversal. However, in
further contrast to Crooks' DFT, we do \emph{not} assume equilibrium steady
states (or detailed balance), any particular starting distribution for the
forward and reverse processes, nor a single heat bath system (or any specific
bath structure). Instead, we require only the functionals $\gWex$ and $\gQhk$ as
defined in Eqs.~\eqref{eq:1L_phi} and \eqref{eq:Qhk_general}, along with an
additional one---the (unitless) \emph{nonsteady-state free energy}:
\begin{align}
\Fnss_\param \pqty{\bm{\mu}, \xx}
  & \defn \ln \frac{\mu\pqty{\xx}}{\pi_\param \pqty{\xx}}
  ~.
\label{eq:Fnss}
\end{align}
Its name derives from its indicating how far a given distribution is from the
associated steady-state distribution. Indeed, on state averaging we have
$\braket{\mu}{\Fnss_\param} = \DKL{\mud}{\bm{\pi}_\param}$, where
$\DKL{\bm{p}}{\bm{q}}$ is the Kullback-Leibler divergence between
distributions $p$ and $q$ \cite{Cove06a}. As with the other
functionals generalized to the stochastic process picture, it carries
meaning---departure from steady-state conditions\\---outside of energetic or
thermal assumptions.

Given this, Ref.~\cite{riechersFluctuationsWhenDriving2017}'s DFT is:
\begin{align}
\frac{
    \RP{x_{N:0}}{\muR}
  }{
    \FP{x_{0:N}}{\muF}
  }
  = \ex^{-\pqty{\gWex + \gQhk - \Delta \Fnss}}
  ~,
\label{eq:NESS_DFT}
\end{align}
where $\Delta \Fnss = \Fnss_{\param_N} \pqty{\muR, \xx_N} - \Fnss_{\param_0}
\pqty{\muF, \xx_0}$ is a correction due to starting the forward and reverse
processes out of steady state. If we began the forward and reverse processes in
their associated steady-state distributions, by definition we would have $\Delta
\Fnss = 0$.

As a mathematical statement involving a stochastic process' trajectories,
their probabilities, and the functionals $\gWex$, $\gQhk$, and $\Delta \Fnss$
we have so far defined, Eq.~\eqref{eq:NESS_DFT} holds \emph{independent} of any
thermodynamic assumptions. Yet, as before, reducing it to thermodynamically
meaningful cases is straightforward and illuminates several important
considerations when moving from ESS into NESS regimes.

\subsection{Multiple NESS IFTs and second laws}

In the ESS case, Eq.~\eqref{eq:NESS_DFT} reduces neatly to Crooks' DFT of
Eq.~\eqref{eq:crooks_DFT}, which under integration directly yields Jarzynski's
equality. However, the NESS setting introduces more freedom under this type of
integration.

Consider rearranging Eq.~\eqref{eq:NESS_DFT} like so:
\begin{align}
  \RP{x_{N:0}}{\muR} = \FP{x_{0:N}}{\muF} \ex^{-\pqty{\gWex + \gQhk - \Delta \Fnss}}
  \textnormal{.}
\label{eq:NESS_DFT_R1side}
\end{align}
Then integrate both sides over \emph{all} trajectories $x_{0:N}$. The righthand
side directly yields the forward trajectory average of the exponential, while
the lefthand side yields $1$ by probability conservation (and since the sets
of all measurable forward and reverse trajectories are the same set). This gives
a generalized IFT:
\begin{align}
  1 = \expval{ \ex^{-\pqty{\gWex + \gQhk - \Delta \Fnss} } }
\label{eq:NESS_IFT_1}
\end{align}
and---via Jensen's inequality---a generalized second law:
\begin{align}
  \expval{\gWex} + \expval{\gQhk} - \Delta \DKL{\bm{\mu}}{\bm{\pi}} \geq 0
  \textnormal{,}
\label{eq:NESS_2L_1}
\end{align}
where $\Delta \DKL{\bm{\mu}}{\bm{\pi}} = \DKL{\muR}{\bm{\pi}_\textnormal{R}} -
\DKL{\muF}{\bm{\pi}_\textnormal{F}}$. This latter term is a classical analogue
to the ``initial-state dependence'' of
Ref.~\cite{riechersInitialstateDependenceThermodynamic2021}; it quantifies
additional entropic dissipation when beginning and ending out of the
steady-state distribution.

Yet Eq.~\eqref{eq:NESS_IFT_1} is not unique. To take one example, by direct
calculation as in Ref.~\cite{hatanoSteadyStateThermodynamicsLangevin2001}, we
find for Markov dynamics:
\begin{align}
  &1 = \expval{\ex^{- \Fnss_{\param_0}\pqty{\bm{\mu}_\textnormal{F}, \, x_0} - \gWex}}
\label{eq:hatanosasa} \\
  &\implies \expval{\gWex} + \DKL{\muF}{\bm{\pi}_\textnormal{F}} \geq 0
  \textnormal{,}
\label{eq:hatanosasa_2L}
\end{align}
a slight generalization of their result with the inclusion of initial-state
dependence (thereby relaxing the requirement of steady-state initial
conditions). In other words, a second law holds for $\expval{\gWex}$ itself,
not just for its sum with $\expval{\gQhk}$. This is not only a meaningfully
different bound, but this IFT also does not result naturally from the
underlying DFT.

To take another example, directly substituting $\expval{\gQhk} =
\expval{\mathcal{Q}} - \expval{\gQex}$ into Eq.~\eqref{eq:NESS_IFT_1}
generalizes to a different IFT:
\begin{align}
  1 = \expval{\ex^{-\pqty{\mathcal{Q} + \Delta \phi - \Delta \Fnss}}}
  \label{eq:NESS_IFT2}
  \textnormal{,}
\end{align}
first proven in Ref.~\cite{speckIntegralFluctuationTheorem2005}.

Finally, taken on their own, Eqs.~\eqref{eq:NESS_2L_1} and
\eqref{eq:hatanosasa_2L} do not imply a third NESS IFT---and so Second
Law---for housekeeping heat alone; cf. again
Ref.~\cite{speckIntegralFluctuationTheorem2005}. However, as we now show, it is
implied rather directly by the combination of Eqs.~\eqref{eq:NESS_DFT} and
\eqref{eq:hatanosasa}.

First rearrange Eq.~\eqref{eq:NESS_DFT} as:
\begin{align*}
  &\RP{x_{N:0}}{\muR} \ex^{\gWex - \Fnss_{\param_N}\pqty{\muR, x_N}} \\
  &\qquad = \FP{x_{0:N}}{\muF} \ex^{- \gQhk - \Fnss_{\param_0}\pqty{\muF, x_0}}
  ~.
\end{align*}
Again, we wish to integrate both sides over all $x_{0:N}$, but at first glance
the lefthand side (first line) poses an issue: $\gWex$ refers to the excess
work over a trajectory driven by the forward protocol, while $\mathcal{R}$ is
the probability of a trajectory as driven by the reverse protocol. 

Fortunately, $\gWex$ is odd under full time reversal: $\gWex =
-\gWex^{\textnormal{R}}$, where $\gWex^{\textnormal{R}}$ is the excess work
generated by the time-reversed trajectory driven by the time-reversed protocol.
This matches up driving protocols and ``initial'' conditions on the lefthand
side. Since Eq.~\eqref{eq:hatanosasa} holds regardless of the chosen protocol or
starting distribution, under integration the lefthand side is unity. The
righthand side, meanwhile, becomes simply the forward trajectory average of its
argument. And, we obtain the generalized IFT for housekeeping heat:
\begin{align}
  &1 = \expval{\ex^{- \Fnss_{\param_0}\pqty{\mu_\textnormal{F}, \, x_0} - \gQhk}}
  \label{eq:IFT_Qhk} \\
  &\implies \expval{\gQhk} + \DKL{\muF}{\bm{\pi}_\textnormal{F}} \geq 0
  \textnormal{,}
  \label{eq:Qhk_2L}
\end{align}
once again extended to include the effects of initial-state dependence. 

To adopt Ref.~\cite{speckIntegralFluctuationTheorem2005}'s language, these
IFTs---for $\gWex + \gQhk$, $\gWex$-only, $\mathcal{Q} + \Delta \phi$, and
$\gQhk$-only---are ``genuinely different'' but no longer require especially
``different derivation[s]'' nor restrictive physical assumptions. To emphasize
the former point, though: just as with the equilibrium second law
$\expval{\Delta S} \geq 0$, these hold only under full trajectory averaging.
That is, individual rare trajectories (or sets thereof) may well produce
negative excess works, negative housekeeping heats, or both.
Sec.~\ref{subsec:results_TCFT} explores these consequences for our
example ion channels.

Notably absent is the notion of steady-state free energy, analogous to the
equilibrium free energy from Jarzynski's equality. Defining one for general
NESS systems remains problematic, in part since the steady-state distributions
may no longer be Boltzmann. Instead, the excess work subsumes what would have
been a steady-state free energy difference, and we work directly with it. The
downside, however, is the inability to extract such a free energy as a ``steady
state'' quantity separate from the path-dependent nonequilibrium dynamical
ones. Indeed, this was an extremely important consequence of
Eq.~\eqref{eq:jarzynski}.

The suite of IFTs given by our Eqs.~\eqref{eq:NESS_2L_1},
\eqref{eq:hatanosasa_2L}, and \eqref{eq:Qhk_2L}, however, do include strong
connections between path-independent and path-dependent quantities in the form
of initial-state dependence and changes in steady-state surprisal. Unlike for
ESS systems, however, even in well-controlled NESS thermal examples applying the
IFTs presents a rather serious experimental challenge: direct measurement of
\emph{heat} (most notably housekeeping heat). Even when testing FTs phrased in
terms of heat, often \emph{work} (excess or not) is experimentally tracked
\cite{trepagnierExperimentalTestHatano2004}. And so, we expect direct
measurement to be a key, requisite step in leveraging the resulting FTs to
analyze experimental NESS systems.

\subsection{NESS trajectory class fluctuation theorem}
\label{subsec:TCFT_theory}

With appropriate DFTs and IFTs for NESS systems now in hand, we are confronted
with yet another challenging experimental tradeoff. Just as the IFTs suffer from
extremely rare-but-large contributions, the DFTs require precise control and
measurement of \emph{individual} realizations, as well as accurate estimations
of individual realization probabilities (or their ratios). This is often
intractable even in principle. For example, as experimental systems, the ion
channels considered shortly do not permit measurement of the conformational
states themselves. Instead, ionic current is the only observable. Moreover, the
state space topology varies with each individual rate model
\cite{vandenbergSodiumChannelGating1991}. This is all to say that thermodynamic
analysis requires a more flexible intermediary between the DFT's
trajectory-level information and the IFT's ensemble-level information.

Ref.~\cite{Wims19a} recently
provided just such an intermediary for ESS systems---the \emph{trajectory
class} FT (TCFT). At root, it relates the forward and reverse probabilities of
an arbitrary subset of trajectories---the \emph{trajectory class} $C$ as
introduced earlier---to the average exponential work \emph{within} that
trajectory class.  In this way, the TCFT is maximally adaptable to experimental
conditions: It need neither suffer rare-event errors nor require
individual-trajectory-level control. Instead, whatever the unique experimental
conditions at hand, it provides a framework for laying out an associated FT. As
a practical matter, the TCFT has already provided a diagnostic tool for
monitoring the thermodynamics of successful and failed microscopic information
processing in superconducting flux logic
\cite{Wims19a,sairaNonequilibriumThermodynamicsErasure2020}.

The following extends Ref.~\cite{Wims19a}'s ESS TCFT (Eq.~(3) there) in two
ways. First, we allow for NESS systems. Second, we allow starting the forward
and reverse processes in arbitrary distributions $\muF$ and $\muR$,
respectively. This results in our exponential NESS TCFT, derived in
App.~\ref{app:ness_tcft}:
\begin{align}
  \frac{
    \RP{C_\textnormal{R}}{\muR}
  }{
    \FP{C}{\muF}
  }
  = \expval{\ex^{-\pqty{\gWex + \gQhk - \Delta \Fnss}}}_C
  ~,
\label{eq:NESS_TCFT}
\end{align}
where $\expval{\cdot}_C$ denotes the conditionally-weighted average over only
those trajectories in the class $C$ and the \emph{reverse trajectory class}
$C_\textnormal{R} \defn \set{x_{N:0} \, \vert \, x_{0:N} \in C}$.

Eq.~\eqref{eq:NESS_TCFT} imports to the NESS setting all the benefits of
the TCFT. Most notably, it adapts readily to a variety of experimental
conditions while maintaining robust statistics. The associated DFT and IFT
emerge simply by setting the class $C$ to be a single trajectory or the set of
all trajectories, respectively. Equation~\eqref{eq:NESS_TCFT}, as with its ESS
counterpart, allows selecting trajectory classes most accessible in a
particular experimental configuration and \emph{then} proposes the appropriate
theory against which to test.

Once again, in this form Eq.~\eqref{eq:NESS_TCFT} makes only two assumptions
about a stochastic process, as outlined previously: a unique stationary
distribution for each $\param$ and an even state space. It reproduces
Ref.~\cite{Wims19a}'s TCFT given ESS assumptions. Similarly, it reproduces
Ref.~\cite{riechersFluctuationsWhenDriving2017}'s Eq.~(52) when the class is
chosen to start and end in a particular desired subset of states. However, our
main result holds independently of any energetic, Markovian, or particular class
assumption.

Generalization to NESS systems is not without caveat, however. $\gQhk$ plays a
central role and we do not have our state- and path-independent equilibrium free
energy to extract from the average and estimate. This suggests experimentally
tracking the housekeeping heat itself is key to understanding nondetailed
balanced, NESS systems. (Alternatively, one could monitor the total heat per
Eq.~\eqref{eq:NESS_2L_1}.) This is not surprising, considering $\Qhk$ is
\emph{the} defining difference between an ESS and NESS system.

\section{Na$^+$ and K$^+$ Ion Channels}
\label{sec:ion_intro}

\begin{figure}[ht]
\centering
\includegraphics[width=\columnwidth]{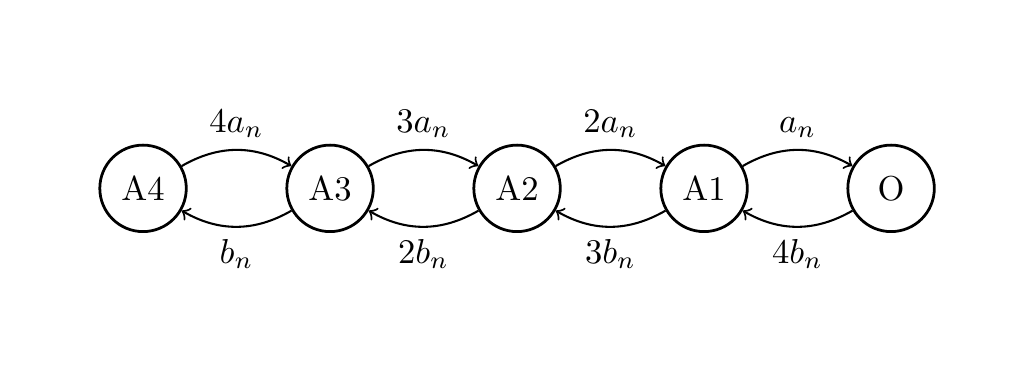}
\caption{Continuous-time Markov chain model of the K$^+$ channel adapted from
	Fig.~5.12 of Ref.~\cite{dayanTheoreticalNeuroscienceComputational2001}.
	Self-transitions are implied. In the states labeled A$n$, a number $n \in
	\{1,2,3,4\}$ activation gates close the channel. O labels the open
	channel state, the only one in which K$^+$ current can flow through the
	channel. The rate parameters $a_n$ and $b_n$ are voltage-dependent; their
	functional forms are given in Eqs.~\eqref{eq:bn}. This channel model is
	fully detailed-balanced, in the sense that Eq.~\eqref{eq:higher_order}
	vanishes for every allowed transition pair.
	}
\label{fig:K_MC}
\end{figure}

\begin{figure}[ht]
\centering
\includegraphics[width=\columnwidth]{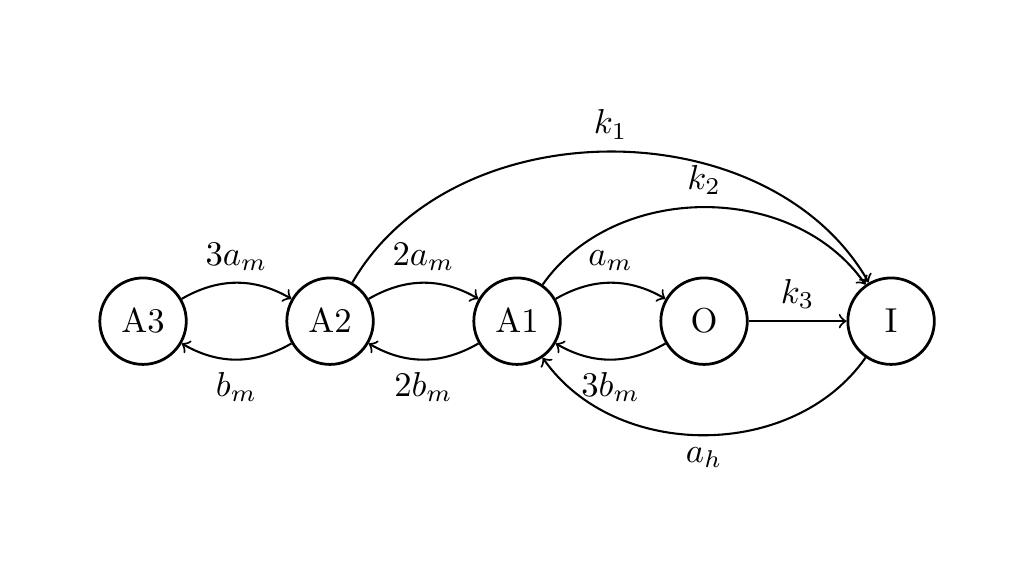}
\caption{Continuous-time Markov chain model of the Na$^+$ channel adapted from
    Fig.~5.13 of Ref.~\cite{dayanTheoreticalNeuroscienceComputational2001}.
    Self-transitions are implied. In the states labeled A$n$, a number $n \in \{1,
	2, 3 \}$
    activation gates close the channel. O labels the open channel state,
	in which Na$^+$ current flows through the channel. Finally, I labels the
	channel's inactivation by its inactivation gate---its so-called ball and
	chain. The rate parameters $a_m$, $b_m$, and $a_h$ are all
	voltage-dependent; their functional forms are given in 
  Eqs.~\eqref{eq:bm}--\eqref{eq:ks}. The rate constants $k_1$, $k_2$, and $k_3$
	are given by Eqs.~\eqref{eq:ks}. Unlike the K$^+$ channel, this model of
	the Na$^+$ channel features one-way transitions in the rate
	dynamic---states O to I and A2 to I. These transitions are maximally
	irreversible. These imply divergent infinitesimal housekeeping heat in the
	sense of Eq.~\eqref{eq:higher_order}. In addition to these, many of the
	other transition pairs do \emph{not} satisfy detailed balance---Eq.~\eqref{eq:higher_order} evaluates finite but nonzero.
    }
\label{fig:Na_MC}
\end{figure}

Armed with this toolkit, we are now ready to probe the thermodynamic
functionality of two example biophysical systems:
Ref.~\cite{dayanTheoreticalNeuroscienceComputational2001}'s delayed-rectifier
potassium (K$^+$) and fast sodium (Na$^+$) voltage-gated ion channels. (See its
Figs.~5.12 and 5.13, reproduced in our Figs.~\ref{fig:K_MC} and \ref{fig:Na_MC},
respectively.) These single-channel models are based on relatively more
macroscopic descriptions of channel ensembles due to Hodgkin and Huxley
\cite{hodgkinQuantitativeDescriptionMembrane1952}. However, they better
represent the interdependencies between molecular-conformational transformations
and more accurately reproduce experimentally-observed currents, especially for
the Na$^+$ channel \cite{dayanTheoreticalNeuroscienceComputational2001}.

The models are both continuous-time Markov chains (CTMCs), whose dynamics are
described by the stochastic master equation:
\begin{align}
  \dv{}{t} \bra{\mu\pqty{t}} = \bra{\mu\pqty{t}} \aGmat{}
\textnormal{.}
\label{eq:master_eq}
\end{align}
The row vector $\bra{\mu\pqty{t}}$ specifies the \emph{state distribution} or
\emph{mixed state} at time $t$; its elements are $\mus\pqty{\xx,t} \defn
\pr{\ST\pqty{t} = \xx}$. The transition rate matrix $\aGmat{}$ is controlled by
the protocol and, thus, varies with time. The would-be steady-state
distributions for each $\param$ are given by:
\begin{align}
\bra{\pi_\param} \mathbf{G}_\param = \bra{0}
  ~,
\label{eq:ssd_defn}
\end{align}  
with $\bra{0}$ the all-$0$ vector.

The transition-rate matrices corresponding to the two channels are:
\begin{widetext}
  \begin{ceqn}
    \begin{align}
      \GK & = 
        \begin{bmatrix}
          -4a_n & 4a_n & 0 & 0 & 0 \\
          b_n & -\pqty{3a_n + b_n}\phantom{+k_1} & 3a_n & 0 & 0 \\
          0 & 2b_n & -\pqty{2a_n + 2b_n}\phantom{+k_2} & 2a_n & 0 \\
          0 & 0 & 3b_n & -\pqty{3b_n + a_n}\phantom{+k_3} & a_n \\
          0 & 0 & 0 & 4b_n & -4b_n
        \end{bmatrix}
      \label{eq:GK} 
      \textnormal{ and} \\
      \GNa & = 
        \begin{bmatrix}
          -3a_m & 3a_m & 0 & 0 & 0 \\
          b_m & -\pqty{2a_m + b_m + k_1} & 2a_m & 0 & k_1 \\
          0 & 2b_m & -\pqty{a_m + 2b_m + k_2} & a_m & k_2 \\
          0 & 0 & 3b_m & -\pqty{3b_m + k_3} & k_3 \\
          0 & 0 & a_h & 0 & -a_h\phantom{4}
        \end{bmatrix}
      \label{eq:GNa}
      \textnormal{.} 
    \end{align}
  \end{ceqn}
Letting $\param$ denote the transmembrane voltage, the associated transition
rates are:
\begin{alignat}{2}
a_m \pqty{\param} & = 
  \frac{
    \pqty{\param + 40\textnormal{ mV}}/10\textnormal{ mV}
  }{
    1 - \exp\pqty{-\pqty{\param + 40\textnormal{ mV}}/10\textnormal{ mV}}
  }
\textnormal{,}
&& b_m \pqty{\param} = 
  4 \exp\pqty{-\pqty{\param + 65\textnormal{ mV}}/18\textnormal{ mV}}
\textnormal{,} 
\label{eq:bm} \\
a_h \pqty{\param} & = \frac{7}{100} 
  \exp\pqty{-\pqty{\param + 65\textnormal{ mV}}/20\textnormal{ mV}}
  \textnormal{,} 
&& k_1 = \frac{6}{25} \textnormal{ ms}^{-1} \textnormal{, }
  k_2 = \frac{2}{5} \textnormal{ ms}^{-1} \textnormal{, }
  k_3 = \frac{3}{2} \textnormal{ ms}^{-1} \textnormal{,} 
\label{eq:ks} \\
a_n \pqty{\param} & = \frac{
    \pqty{\param + 55 \textnormal{ mV}}/100\textnormal{ mV}
  }{
    1 - \exp\pqty{-\pqty{\param + 55\textnormal{ mV}}/10\textnormal{ mV}}
  }
  \textnormal{, and} \qquad
&& b_n \pqty{\param} = \frac{1}{8} 
  \exp\pqty{-\pqty{\param + 65\textnormal{ mV}}/80\textnormal{ mV}}
  \textnormal{.} 
\label{eq:bn} 
\end{alignat}
\end{widetext}

We map these CTMC systems to discrete-time stochastic processes by taking
$\param\pqty{t}$ fixed for sufficiently small time intervals $\Delta t$,
generating the transition matrices:
\begin{align}
\mathbf{T}_\param^{\Delta t} \defn \ex^{\Delta t \, \mathbf{G}_\param}
\end{align}
for each such time interval. Having discretized time in this way, they are
examples of the thermodynamic scheme in Fig.~\ref{fig:thermo_visual}, being
surrounded by a single thermal environment at body temperature.

A voltage-gated ion channel's basic function is to selectively allow ions to
permeate a cell membrane. The selection is based on the transmembrane
voltage---the voltage difference between the membrane's inside and outside. In
our models, this difference is specified by the parameter $\param$, and so a
neuronal action potential spike is a specific protocol. Ref.~\cite{dayanTheoreticalNeuroscienceComputational2001}'s K$^+$ and Na$^{+}$
models correspond to channels that play crucial roles in generating and
propagating such spikes in mammalian neuronal axons. Both Markov chain models are
estimated from single-channel experiments.

We selected these two channel models for several reasons.

First, in terms of their biological function, they are comparable: They
accomplish similar tasks, are connected to the same environmental parameters,
and are suitably mesoscopic. That is, despite being more detailed than the
Hodgkin-Huxley ensemble models, neither model accounts for the many additional
molecular degrees of freedom involved in the channel dynamics, be it
steady-state or transient functions. The small effective state spaces in the
Markov chain models reflect this.

One consequence of this implied coarse graining is that any total entropy
production is a lower bound \cite{espositoStochasticThermodynamicsCoarse2012}.
Still, we are able to make headway analyzing their nonequilibrium dynamics
\emph{without} knowledge of the underlying coarse-graining methods---knowledge
missing for the vast majority of mesoscopic complex systems.

Second, the Na$^+$ channel's transition rates do not, in general, satisfy
detailed balance, while the K$^+$ channel's do. Indeed, the Na$^+$ channel model
includes both finitely nondetailed-balanced transition pairs and one-way-only
transition rates, which imply divergent infinitesimal-time housekeeping heat.
These violations of ideality are typical and widely encountered in molecular
biophysical systems, as well as in real-world thermodynamic processes.

To appreciate these nonidealities, note that under our time discretization
and the Markov property, the \emph{infinitesimal}(-time) \emph{housekeeping
heat}---a single ``step'' of Eq.~\eqref{eq:Qhk_general}---for a transition
between states indexed by $i$ to $j$:
\begin{align}
  \bqty{\dd \gQhk}_{ij} &= \lim_{\Delta t \to 0}
  \ln \frac{\ssda{x_i} \Taij{ij}}{\ssda{x_j} \Taij{ji}} 
  \notag \\
  &= \lim_{\Delta t \to 0} \ln 
    \frac{
      \ssda{x_i}
      \bqty{
        \Gmat_\param +\mathcal{O}\pqty{\Delta t}
      }_{ij}
    }{
      \ssda{x_j}
      \bqty{
        \Gmat_\param + \mathcal{O}\pqty{\Delta t}
      }_{ji}
    }
    \textnormal{.}
\label{eq:higher_order}
\end{align}
For systems with one-way-only transition rates, such as from the second to the
fifth state of the Na$^+$ channel (indexing the states left to right, Fig.
\ref{fig:Na_MC}), infinitesimal housekeeping heat diverges. This contrast
between the two channels---wherein one of them exhibits equilibrium steady
states and the other nonequilbrium steady states---allows showcasing several
features of the NESS TCFT and of the NESS framework more broadly. These, in
turn, reveal the dynamical interplay of different modes of thermodynamic
transformation.

As a test case, the K$^+$ channel should satisfy the ESS TCFT (where $\gQhk =
0$) while the Na$^+$ channel should violate it. Both, however, should satisfy
our NESS TCFT of Eq.~\eqref{eq:NESS_TCFT}.

One benefit of the TCFT's averaging over arbitrary trajectory classes comes
from avoiding the divergences implied by one-way transition rates: We select
only those trajectories that do \emph{not} include one-way transitions in the
Na$^+$ channel, but still satisfy the appropriate DFT (and therefore TCFT) with
those trajectories.

In this way, the NESS TCFT allows monitoring nonequilibrium fluctuations in
systems with drastically different steady-state characteristics: detailed
balance on the one hand and spurious divergences on the other.

Yet separating heat into excess and housekeeping components \emph{also} enables
direct comparison of the channels' \emph{adaptive} energetics. Given the same
environmental drive, which components of their dissipations are due solely to
their internal adaptation to that drive? The excess heat, $\gQex$. This remains
true without regard for the divergence implied by one model's steady states. In
essence, we cleave the housekeeping infinity to directly compare adaptive
energetics.

Finally, both models are simple and illustrative. There are many more-detailed
candidate state-space models for the Na$^+$ channel: take those found in
Refs.~\cite{vandenbergSodiumChannelGating1991} and
\cite{palDynamicalCharacterizationInactivation2016}, for instance, whose
variations have important implications for understanding responses to drug
treatments \cite{marbanStructureFunctionVoltagegated1998}. While we do not
analyze them directly, our techniques generalize to \emph{any} such candidate
models straightforwardly and provide an alternative formulation to that of
Ref.~\cite{palDynamicalCharacterizationInactivation2016}. Indeed, our ability to
carry out these thermodynamic analyses provides new grounds for model selection,
contingent on measurement techniques to experimentally extract the appropriate
quantities.

\section{Methods and Results}
\label{sec:results}

Our goal, ultimately, is to describe the nonequilibrium thermodynamics of
driven mesoscopic NESS systems---how they respond thermodynamically to environmental stimuli and attempt to maintain stability. We take up the
challenges here in two ways.

First, Sec.~\ref{subsec:results_TCFT} samples individual trajectories from
both channels under the neurobiologically-plausible action potential \emph{spike
protocol}. Trajectories in hand, it compares $\Qhk$ versus $\Wex$ for each,
showcasing the need for corrected NESS D/TCFTs, revealing various modes of
second-law-type violations allowed of each channel, and discussing the
surprising biophysical functionality these violations imply.

We derived the spike protocol by solving the reduced ODEs (8.5) and (8.6) of
Izhikevich \cite{izhikevichDynamicalSystemsNeuroscience2006} (also presented
earlier \cite{izhikevichSimpleModelSpiking2003}), integrating via the explicit
forward Euler method. We adopt the ``regular spiking'' parameters of their Fig.
8.12, except that we set the membrane capacitance to $1$ pF and input DC pulse
to $80$ pA to change the time scale of a single pulse to $2$ ms, more
accurately reflecting measurements in
Ref.~\cite{dayanTheoreticalNeuroscienceComputational2001}. The protocol begins
with the transmembrane voltage at its resting potential (of $-60$ mV in this
parameter set) and the recovery variable $u$ at $0$. We took $200\,001$
equidistant time steps, resulting in $10$ ns increments.

Second, Sec.~\ref{subsec:avg_excess} calculates the full trajectory-averaged
excess heat and work---$\expval{\Qex}$ and $\expval{\Wex}$, respectively---of
the two channel models under both our spike protocol and the $12$ ms \emph{pulse
protocol} matching Ref.~\cite{riechersFluctuationsWhenDriving2017}'s and
providing for direct comparison with their results. We took the same number of
equidistant time steps, resulting in $60$ ns increments. In the spike case, we
directly compare for the first time the detailed \emph{adaptive} energetics of
the two channel types under a neurobiologically-plausible protocol. Our analysis
both reveals functionality not visible under a pulse drive and highlights the
preceding theoretical framework's ability to directly compare the channels'
adaptive energetic response to the same drive, despite their dramatically
different steady-state behaviors.

The ion channels are examples of Fig.~\ref{fig:thermo_visual}'s scheme with a
single heat bath, so we have that $\gQhk \to \beta \Qhk$, $\gWex \to \beta
\Wex$, and $\gQex \to \beta \Qex$. For convenience, then, we label all
thermodynamic axes in units of $\bqty{\kb T}$. In more general settings,
however, these functionals are purely dynamical quantities, to be understood and
interpreted as indicated in Secs.~\ref{subsec:core_definitions} and
\ref{subsec:db_housekeeping}.

Admittedly, the selected ion channel models are not realistic in the sense that
they do not incorporate feedback between the transmembrane potential and the
ion channel states themselves. (Or, put another way, they ignore correlation
between channels.) This feedback is crucial to \emph{in vivo} generation of the
spike patterns. In one sense, this simplification is actually an
\emph{advantage} of our approach, since we ask: \emph{Given} a particular
transmembrane protocol---regardless of how it got there---how do these
individual channels respond? How do they absorb and dissipate energy in
response to this environment?

\subsection{NESS TCFT reveals thermal response}
\label{subsec:results_TCFT}

This section compares the detailed-balanced dynamics of the K$^+$ channel
with the nondetailed-balanced dynamics of the Na$^+$ channel. It demonstrates
agreement between ESS and NESS FTs in the former, but violation in the latter.
This exposes the channels' different dynamical responses---how thermodynamic
fluxes of energy and entropy support their distinct biophysical functioning. The
TCFT's flexibility allows us to select only trajectories-of-interest and take
partial sums on either side of the underlying DFT. This helps not only to gather
experimental statistics---improving statistical efficiency---but also to
generate statistics from models, as the following does.

While Eq.~\eqref{eq:higher_order}'s first-order approximation is valid in the
infinitesimal time limit, \emph{any} finite time step---no matter how
small---maps every zero in the transition-rate matrices to nonzero values in the
discrete-time transition matrices. As long as any state can transition to any
other \emph{eventually} in the rate dynamic, we observe a \emph{direct}
transition from any state to any other state after \emph{any finite time}.
Mathematically, this results from higher-order terms in the matrix
exponentials.

Since we wish to explicitly highlight the differences between the channels---the
ESS in the K$^+$ case and the divergent transitions in the Na$^+$---we take the
first-order approximation of Eq.~\eqref{eq:higher_order}. Formally, it defines a
distinct discrete-time dynamic compared to taking the full matrix exponentials,
but the fluctuation theorems apply just as well to this approximated dynamic. In
sampling trajectories, we avoid the divergent Na$^+$ transitions altogether by
selecting only paths that do not include them, yet another advantage the TCFT
affords. This does not alter the TCFT's validity as long as we accurately
collect the probabilities of the selected trajectories.

We \emph{can} collect those probabilities, having the full transition dynamic in
hand. However, simulating $200\,001$-step trajectories, the resulting
probabilities are extraordinarily small. To ameliorate numerical precision
issues, we instead directly collect the natural logarithms of trajectory
probabilities. Finally, since we wish to isolate the differences between the
channels due to NESSs (or, equivalently in our case, to nondetailed-balanced
dynamics), we make one last simplifying assumption before numerical simulation:
We begin all forward and reverse processes in their local stationary
distributions, setting $\Delta \Fnss = 0$.

This simplifies Eq.~\eqref{eq:NESS_TCFT}'s DFT kernel to:
\begin{align}
  \ln \frac{\FP{x_{0:N}}{\bm{\pi}_\textnormal{F}}}{\RP{x_{N:0}}{\bm{\pi}_\textnormal{R}}} = 
     \gWex\bqty{x_{0:N}} + \gQhk\bqty{x_{0:N}}
  ~.
\label{eq:log_NESS_DFT_begin-SS}
\end{align}

\begin{figure*}[ht]
\centering
\includegraphics{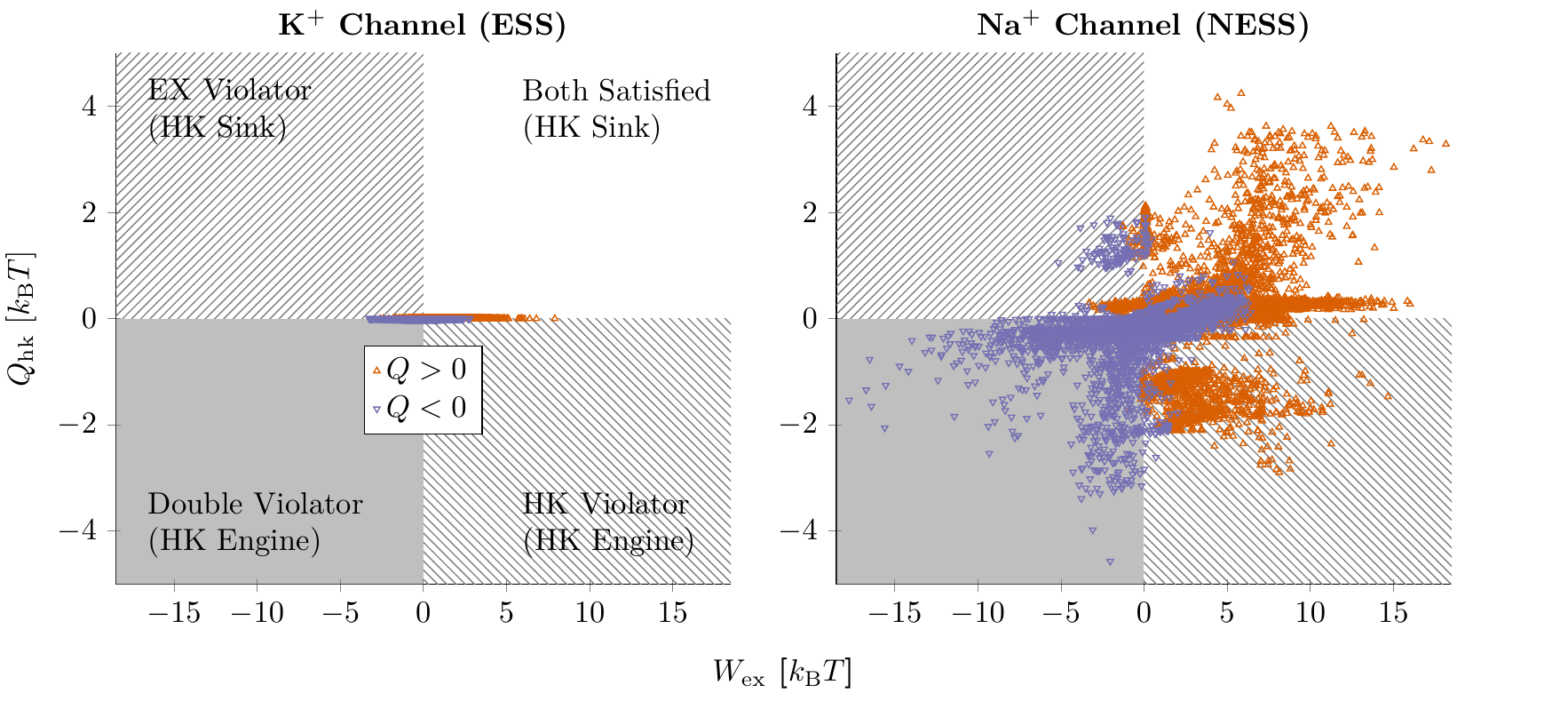}
\caption{The two channel models compared via their nonequilibrium excess work
    and housekeeping heat distributions, respectively, in response to the spike
    protocol drive. Nonzero values of $\Qhk$ indicate violations of the Crooks
    DFT Eq.~\eqref{eq:crooks_DFT}, where the corrected NESS DFT
    Eq.~\eqref{eq:NESS_DFT} is needed. In addition, the axes here are of
    quantities with associated independent second laws; see
    Eqs.~\eqref{eq:hatanosasa_2L} and \eqref{eq:Qhk_2L}. And so, the labeled
    quadrants carry thermodynamic (and, in this case, biophysical) meaning as
    single-shot violations of each statistical second law.
	}
\label{fig:combined_DFT}
\end{figure*}

Comparing this to Crooks' DFT as in Eq.~\eqref{eq:crooks_DFT} reveals the
presence of $\gQhk$ as the only difference. For an ESS system, this should
vanish for all inputs; otherwise, it represents a violation of Crooks' DFT by a
factor of $\ex^{\gQhk}$. To probe the violation, Fig.~\ref{fig:combined_DFT}
directly plots $\Qhk$---via Eq.~\eqref{eq:Qhk_general}---on the vertical axis,
where each point represents an individual trajectory. We plot these values
against $\Wex$---with the discrete form of Eq.~\eqref{eq:1L_phi}---on the
horizontal axis to aid interpretation: Via Eqs.~\eqref{eq:hatanosasa_2L} and
\eqref{eq:Qhk_2L}, there are individual second laws for both the generalized
housekeeping heat and the generalized excess work. As with the familiar
equilibrium second law, however, these are strictly true only on full trajectory
averaging.

To arrive at Fig.~\ref{fig:combined_DFT}, we sampled trajectories according to
their distributions as given by each channel's first-order dynamics under spike
driving. For the Na$^+$ channel, as previously mentioned, this excludes the
one-way-only transitions. For the K$^+$ channel, we obtained $9\,626$ individual
trajectories; for the Na$^+$ channel, we obtained $23\,834$.

Plotting housekeeping heat against excess work in Fig.~\ref{fig:combined_DFT}
directly visualizes the independent kinds of negative entropy trajectories:
where $\gWex < 0$, we have single-shot violations of the familiar second law. In
the isothermal environment of the ion channel models, trajectories for which
this is the case imply channels that, under the spike protocol's drive, funnel
energy to the work reservoir. Where $\gQhk < 0$, however, we have a new kind of
second law violation unique to the NESS setting: in context, these are channels
which have taken energy \emph{from} the heat bath to maintain NESS conditions,
rather than dissipated to it. For this reason, in Fig.~\ref{fig:combined_DFT},
we label these quadrants ``housekeeping (HK) engines.'' Where total heat is also
negative, the channel as a whole functions as a ``total heat engine,''
but---notably---these possibilities are independent of one and other. To take
but one example: the negative total heat trajectories of the first quadrant act
as heat engines, yet the \emph{housekeeping} part of their total heat remains
dissipative.

As Fig.~\ref{fig:combined_DFT} shows, the additional dimension of thermodynamic
behavior afforded by nonzero housekeeping heat and its associated second law
gives rise to a number of otherwise inaccessible combinations. Driving the
channels according to the biologically-plausible spike protocol also reveals a
greater range of possible Crooks DFT violations than did the more artificial
pulse-driven result of Ref.~\cite{riechersFluctuationsWhenDriving2017}, with
only several violations. Taken together, Fig.~\ref{fig:combined_DFT} reveals a
rich taxonomy of thermodynamic behaviors for the Na$^+$ channel---behaviors that
are not reflected (indeed, flattened) in the K$^+$ channel or, indeed, in
\emph{any} ESS system, where Crooks' DFT is satisfied and $\gQhk = 0$. In
particular, there are four functionally-distinct thermodynamic quadrants,
corresponding to the positive and negative values of $\gQhk$ and $\gWex$, and
labeled by their excess and housekeeping functionality on the K$^+$ channel
plot for clarity.

To be clear, each point on these plots corresponds to a \emph{single}
trajectory-reverse pair that itself is a valid trajectory class. Yet, (i) the
samples themselves may be taken from a special class---for the Na$^+$ channel we
explicitly exclude resource-divergent trajectories---and (ii) any subsample on
the plot corresponds to its own valid trajectory class as well.

The clustering of realized $\Qhk$ in Fig.~\ref{fig:combined_DFT} reveals
additional \emph{structure} in Crooks DFT violations not previously observed.
Apparently, there are distinct thermodynamic mechanisms that generate the
violations. These result directly from the relative frequencies of transitions
as functions of the driving protocol.

\begin{figure}[ht]
  \centering
  \includegraphics[width=\columnwidth]{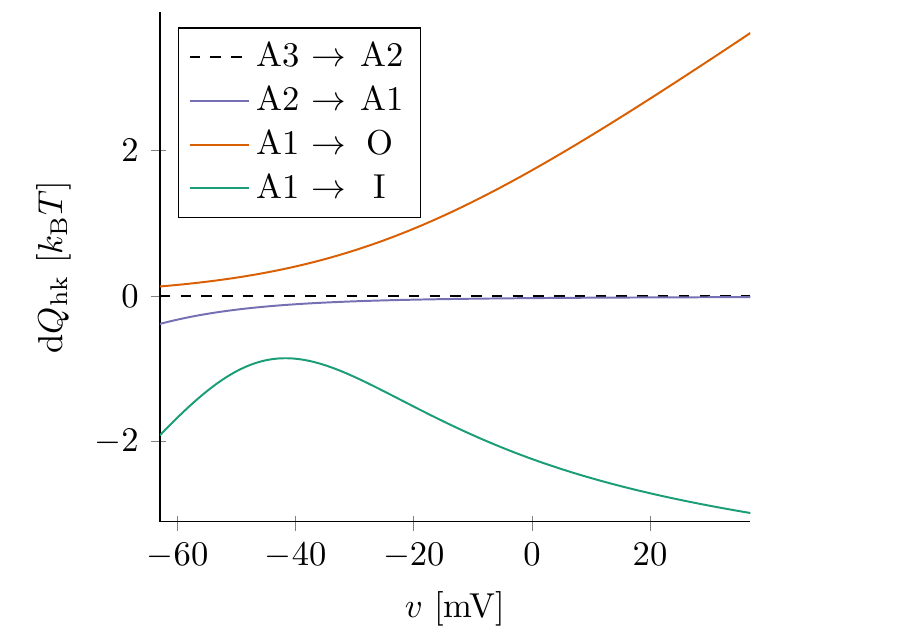}
  \caption{Na$^+$ channel transition rates of housekeeping heat $\Qhk$ as
    functions of transmembrane potential. The total housekeeping heat along any
    stochastic trajectory, driven by any protocol, is the sum of these values
    with the associated state transition-protocol parameter pairs here.
    }
  \label{fig:Na_dQhk}
\end{figure}

To lend additional insight into this structure, Fig.~\ref{fig:Na_dQhk} plots
the one-step rates of $\Qhk$ production for each allowed transition in our
modified Na$^+$ channel. The A3 $\leftrightarrow$ A2 transition pair is the
only one of this dynamic that is fully detailed-balanced for all inputs; the A2
$\leftrightarrow$ A1 pair is \emph{nearly} detailed-balanced, with very small
housekeeping heat production. By comparison, both of the remaining transitions
\emph{strongly} violate detailed balance, and so contribute the bulk of the
nonzero $\Qhk$.

Physically, the latter two correspond to transitions directly to/from the open
and inactivated channel states. Interestingly, their violations run in opposite
directions. On the one hand, the A1 $\rightarrow$ O transition dissipates
housekeeping heat to the thermal environment---indeed, more as the membrane
voltage rises. On the other hand, the A1 $\rightarrow$ I transition describes
the ``ball and chain'' that plugs the channel without opening, leaving no
opportunity for Na$^+$ current to flow. Thermodynamically, this transition
actually \emph{absorbs} housekeeping heat from the thermal environment. Since
the housekeeping heat production rate is odd under transition reversal, these
roles are reversed for the reverse transitions. Thus, the results shown in
Fig.~\ref{fig:combined_DFT} arise directly from integrating those in
Fig.~\ref{fig:Na_dQhk} according to each trajectory-protocol pair.

As a final consideration, we note that the ESS TCFT (and so the Crooks DFT)
\emph{do not claim} to be valid for NESS systems. That said, our results
visually verify the facts that the NESS generalization both extends the range of
validity of the TCFT and reduces in the correct way for ESS systems.

That we have a trajectory class form for the NESS TCFT, captured in our
Eq.~\eqref{eq:NESS_TCFT}, imports its ESS progenitor's flexibility. That is, we
need capture neither individual trajectory-level information to verify the DFT
nor accurately sample the full trajectory space for an IFT.

That said, experimental verification remains a significant challenge.
Generalizing to NESS systems requires not only the excess work distribution but
housekeeping heats as well. Indeed, these results suggest that carefully
considering how to measure housekeeping dissipation is crucial to
characterizing fluctuations in NESS systems. As Fig.~\ref{fig:combined_DFT}
demonstrates, improper accounting leads in general to TCFT \emph{violations}
and, if the ESS FTs are used to estimate free energy differences, to
potentially drastically mischaracterizing the system of interest.

\subsection{Average excess energetics}
\label{subsec:avg_excess}

Despite the channels' distinct thermodynamic functioning as revealed by the
TCFT, we can compare the channels' adaptive energetics via the excess works and
heats. We begin by directly calculating the full trajectory averages, obtaining
for discrete time:
\begin{align}
  -\expval{\gQex} &= \sum_{n=0}^{N-1} \braket{\mu\pqty{t_{n+1}} 
  - \mu\pqty{t_{n}}}{\phi_{\param_n}}
  \textnormal{ and} \label{eq:avg_Qex}\\
  \expval{\gWex} &= \sum_{n=0}^{N-1} \braket{\mu\pqty{t_n}}
  {\phi_{\param_{n+1}} - \phi_{\param_n}}
\textnormal{,}
\label{eq:avg_Wex}
\end{align}
in agreement with Ref.~\cite{riechersFluctuationsWhenDriving2017}. As above, we
set the initial distributions to the local stationary distribution for
convenience. Armed with the discrete protocols, time steps, and starting
distributions, we directly evaluate the mixed states [Eq.~\eqref{eq:master_eq}]
and steady-state distributions [Eq.~\eqref{eq:ssd_defn}] for each time step.
These are all that is needed to calculate $\expval{\gQex}$ and $\expval{\gWex}$
via Eqs.~\eqref{eq:avg_Qex} and \eqref{eq:avg_Wex}.

\begin{figure}[ht]
\centering
\includegraphics[width=\columnwidth]{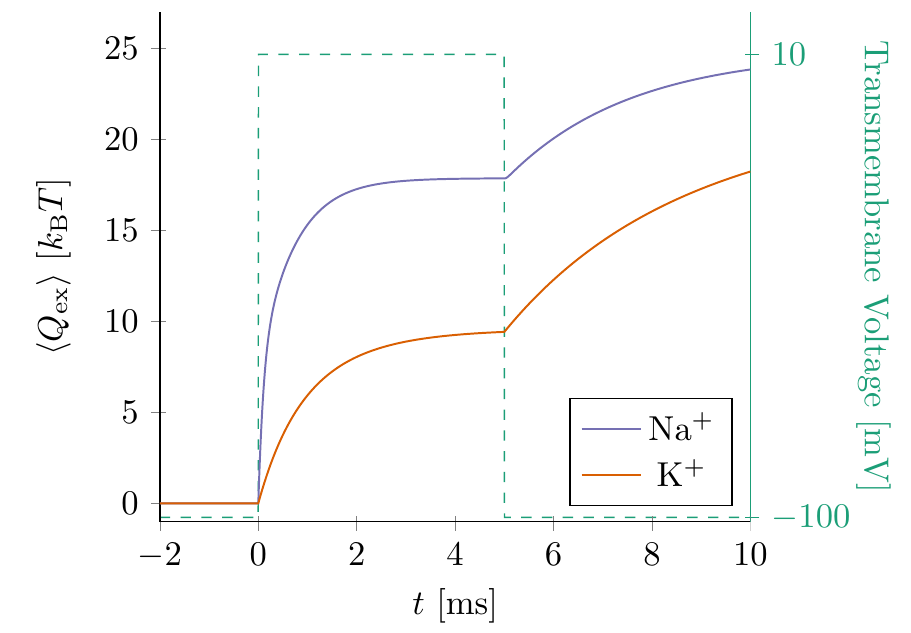}
\caption{Excess heats (solid lines) for both channels under the pulse protocol
	(dashed line). The K$^+$ channel is less dissipative. Both expend energy as
	they relax to environmentally-induced steady states.
	}
\label{fig:qex_both_pulse}
\end{figure}

\begin{figure}[ht]
\centering
\includegraphics[width=\columnwidth]{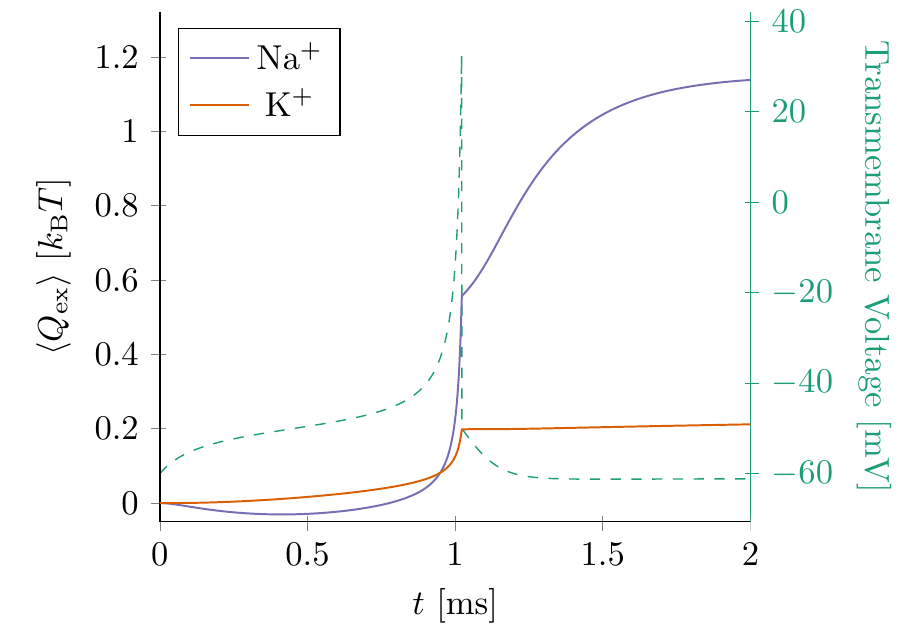}
\caption{Excess heats (solid lines) for both channels driven by the spike
	protocol (dashed line). Under this more biologically realistic protocol, the
	Na$^+$ dissipates significantly more than the K$^+$ channel and does so
	responding much more rapidly to changes in membrane voltage. This suggests a
	tradeoff between the speed of the channel's response and its dissipation,
	one not necessarily present in the more artificial pulse protocol.
	}
\label{fig:qex_both_spike}
\end{figure}

Figs.~\ref{fig:qex_both_pulse} and \ref{fig:qex_both_spike} give the
simulation results for excess heat.

First, driven by the pulse, the K$^+$ channel dissipates less excess heat over
the course of this protocol. Its rate of relaxation to steady
state---corresponding to constant epochs in the protocol---appears slower than
the Na$^+$ channel's on the jump from $-100$ to $10$ mV, but faster on the
subsequent drop back down to $-100$ mV.

The spike protocol paints a very different picture. Here, while the Na$^+$
channel still dissipates (in this case, significantly) more over the course of
the protocol, it \emph{also} appears to respond much more quickly to changes in
the protocol than does the K$^+$ channel. A tradeoff appears: the cost of the
Na$^+$ channel adapting more \emph{quickly} to its environment is that it
dissipates more in the process. This did not arise when driven by the pulse
protocol. (Likely, this is due to that protocol operating outside of the
``normal'' voltage range for these channels---by dropping as low as $-100$ mV.)

Besides showcasing a detailed energetic comparison between different channels,
the discrepancy between the pulse- and spike-driven behaviors demonstrate that
\emph{in vivo} thermodynamic response can qualitatively differ from that
elicited by voltage-clamp experiments.

\begin{figure}[ht!]
\centering
\includegraphics[width=\columnwidth]{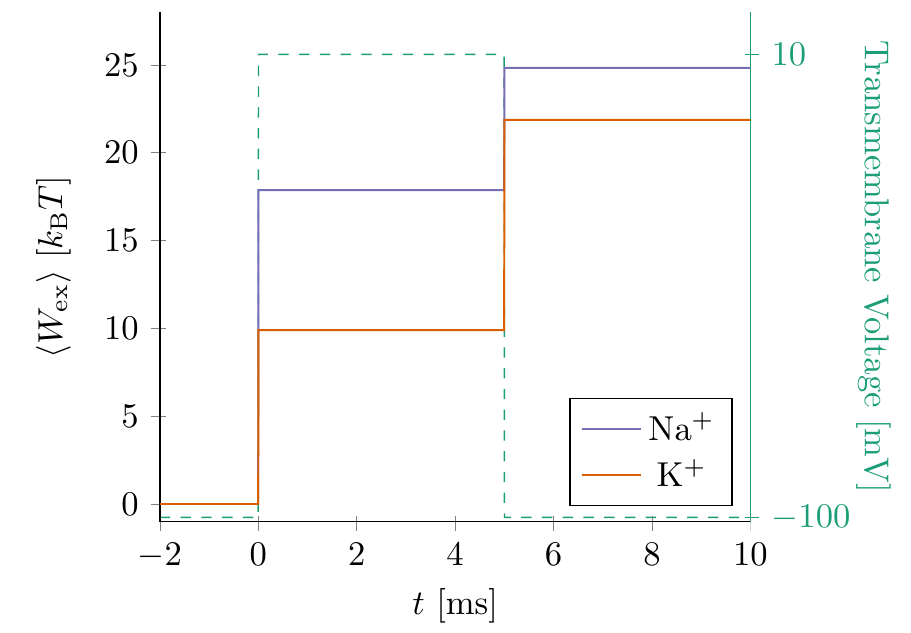}
\caption{Excess works (solid lines) for both channels under the pulse protocol
	(dashed line). Unlike excess heat, excess work is only done upon \emph{change}
	in the driving parameter. Thus, we see changes only at the pulse's rise and
	fall. Much more excess work is done on the Na$^+$ channel than on the K$^+$
	during the rise of the pulse, but these roles are reversed on its fall. Over
	the entire protocol, the Na$^+$ channel produces (slightly) more excess
	environmental entropy.
	}
\label{fig:wex_both_pulse}
\end{figure}

\begin{figure}[ht!]
\centering
\includegraphics[width=\columnwidth]{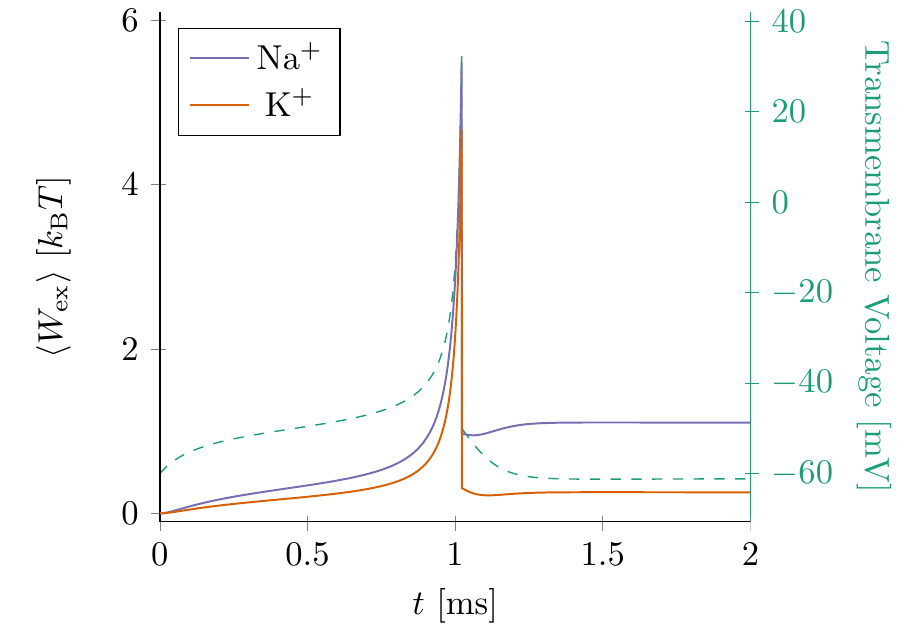}
\caption{Excess works (solid lines) for both channels under the spike protocol
	(dashed line). We see that more excess work is done on the Na$^+$ channel
	across the board. This corresponds to larger environmental entropy production
	and indicates a greater potential for dissipated work in the Na$^+$ channel.
	However, we also see that up until the peak and reset phases of the action
	potential spike, they track very closely before diverging.
	}
\label{fig:wex_both_spike}
\end{figure}

The corresponding results for excess work are given in Figs.
\ref{fig:wex_both_pulse} and \ref{fig:wex_both_spike}. Unlike excess heat, the
excess work is not sensitive to the timescales of each channel's relaxation to
steady state. Instead, it tracks \textit{environmental} entropy produced by the
external drive. Yet it is still sensitive to the dynamics of the individual
channel [per Eq.~\eqref{eq:1L_phi}], and this sensitivity is reflected in the
thermodynamic responses.

As in the excess heat calculations, there is a difference in behavior between
the pulse and spike protocols. In the former, the K$^+$ channel and Na$^+$
channels trade off under the rise and fall of the pulse. Driven by the spike
protocol, however, the Na$^+$ channel induces more environmental entropy
production across the board, though they track extremely closely until the peak
and reset phases of the action potential spike. This reflects not only the
larger potential for dissipation in the Na$^+$ channel under the spike protocol,
but highlights where during the protocol most of the difference arises.

To reiterate, while excess heat is an energetic signature of relaxation to
steady state, these calculations do not assume the system ever \textit{reaches}
such a steady state. While the pulse protocol allows each of the channels to do
so, over the course of the spike we see a \emph{dynamic} dissipation in the two
channels---this is energy expended while attempting to reach an ever-evolving
steady-state target. We close by highlighting that our theoretical developments
enabled quantitatively comparing the channels' adaptive energetics under
realistic environmental stimuli---captured in our
Fig.~\ref{fig:qex_both_spike}---\emph{despite} the departures in their
underlying steady-state dissipation.

\section{Conclusion}
\label{sec:conclusion}

We reviewed and extended the techniques of stochastic thermodynamics,
culminating in a TCFT for NESSs, and even, for nonthermal stochastic processes.
Using these, we analyzed the adaptive and homeostatic energetic signatures of
two neurobiological systems---systems key to propagating action potentials in
mammalian neurons.  Along the way, we developed a toolkit for probing the
nonequilbrium thermodynamics in a broad range of mesoscopic complex systems that
requires little in the way of restrictive assumptions.

Our results exposed a new quantitative structure in how systems appear to
violate equilibrium steady-state assumptions, both warning against and
elucidating the consequences of inappropriately assuming detailed-balanced
dynamics. In this, they suggest a need for both our corrected and generalized
TCFT \emph{and} new experimental tools.

Specifically, for nonequilibrium steady-state systems, tracking housekeeping
entropy production is \emph{crucial} to extracting functionally relevant
thermodynamics and observing an additional \emph{kind} of second law dynamics.
While experimental tests have verified Eq.~\eqref{eq:hatanosasa_2L}, they do not
allow for observing the housekeeping thermodynamics, which play an important and
independent part in assessing a system's functionality.

Our simulations of the averaged excess energetics, in contrast, show how to
compare specific aspects of a system's functionality---the adaptive
energetics---despite what are in this case infinite differences in the steady
state behavior. In essence, these tools allow us to ``cleave off'' the
divergence in the Na$^+$ model's housekeeping heat and still compare the
channels' adaptation to environmental drive on equal footing. Finally, the
spike protocol simulations also identified what \emph{would not} be observed in
traditional patch-clamp experiments on ion channels, namely detailed
differences in response to each segment of an action potential.

Taken together, our development and associated numerical experiments revealed a
rich---and, indeed, necessary---set of tools with which to probe the
nonequilibrium dynamics of mesoscopic complex systems.

\section*{Acknowledgments}
\label{sec:acknowledgments}

The authors thank Paul Riechers, Gregory Wimsatt, Samuel Loomis, Alex Jurgens,
Alec Boyd, Kyle Ray, Adam Rupe, Ariadna Venegas-Li, David Geier, and Adam Kunesh
for helpful discussions; as well as the Telluride Science Research Center for
hospitality during visits and the participants of the Information Engines
Workshops there for their feedback and discussion. This material is based upon
work supported by, or in part by, Foundational Questions Institute Grant
No.~FQXi-RFP-IPW-1902 and the U.S. Army Research Laboratory and U.S. Army
Research Office under Grants No.~W911NF-18-1-0028 and No.~W911NF-21-1-0048.

\appendix

\section{NESS TCFT Derivation}
\label{app:ness_tcft}

In addition to requiring a unique stationary distribution for each protocol
value, we assume that for any $x_{0:N} \in \mathcal{X}^{N+1}$:
\begin{enumerate}
\setlength{\topsep}{-2pt}
\setlength{\itemsep}{-2pt}
\setlength{\parsep}{-2pt}
\item $x_{N:0} \in \xsp^{N+1}$,
\item $\mathcal{P}_{\mud_\textnormal{F}}\pqty{x_{0:N}} \neq 0 \implies
	\mathcal{P}_{\mud_\textnormal{F}}\pqty{x_{N:0}} \neq 0$, and
\item $\mathcal{R}_{\mud_\textnormal{R}}\pqty{x_{N:0}} \neq 0 \implies
	\mathcal{R}_{\mud_\textnormal{R}}\pqty{x_{0:N}} \neq 0$.
\end{enumerate} 
The second and third requirements, in particular, forbid one-way-only
transitions in the discrete-time dynamic. Once we derive the TCFT, we will
discuss the edge cases of completely irreversible trajectories.

Given the preceding constraints, a slightly rearranged form of
Ref.~\cite{riechersFluctuationsWhenDriving2017}'s NESS DFT reads:
\begin{align*}
  \mathcal{R}_{\mud_\textnormal{R}}\pqty{x_{N:0}} 
  = \mathcal{P}_{\mud_\textnormal{F}}\pqty{x_{0:N}} \, 
  \ex^{-\pqty{\gWex + \gQhk - \Delta \Fnss}}
  ~.
\end{align*}
We wish to integrate both sides over a \emph{trajectory class}---the measurable
subset $C \subseteq \xsp^{N+1}$ of trajectories. We also define the
\emph{reverse trajectory class} $C_\textnormal{R} \defn \set{x_{N:0} \, \vert \,
x_{0:N} \in C}$. The following derivation mimics that of Ref.~\cite{Wims19a}
after their Eq.~(F3).

Integrating the lefthand side gives:
\begin{align*}
  \int & \bqty{x_{0:N} \in C} \, \mathcal{R}_{\mud_\textnormal{R}}
  \pqty{x_{N:0}} \dd{x_{0:N}} \notag \\
  &= \int \bqty{x_{0:N} \in C} \, \mathcal{R}_{\mud_\textnormal{R}}
  \pqty{x_{N:0}} \dd{x_{N:0}} \notag \\
  &= \int \bqty{x_{N:0} \in C_\textnormal{R}} 
  \, \mathcal{R}_{\mud_\textnormal{R}}
  \pqty{x_{N:0}} \dd{x_{N:0}} \notag \\
  &= \mathcal{R}_{\mud_\textnormal{R}} \pqty{C_\textnormal{R}}
  ~,
\end{align*}
where $[\cdot]$ is the Iverson bracket.

Integrating the righthand side gives:
\begin{align*}
\int & \bqty{x_{0:N} \in C} \, 
  \mathcal{P}_{\mud_\textnormal{F}}
  \pqty{x_{0:N}} \,
  \ex^{-\pqty{\gWex + \gQhk - \Delta \Fnss}} \dd{x_{0:N}} 
  \notag \\
  &= \int \mathcal{P}_{\mud_\textnormal{F}}
  \pqty{x_{0:N} \cap C} \,
  \ex^{-\pqty{\gWex + \gQhk - \Delta \Fnss}} \dd{x_{0:N}} 
  \notag \\
  & = \mathcal{P}_{\mud_\textnormal{F}}\pqty{C} \times \\
  & \qquad \hspace{-0.15em}\int \mathcal{P}_{\mud_\textnormal{F}}
  \left( x_{0:N} \,\vert\, C \right) \,
  \ex^{-\pqty{\gWex + \gQhk - \Delta \Fnss}} \dd{x_{0:N}}
  \notag \\
  & = \mathcal{P}_{\mud_\textnormal{F}}\pqty{C}
  \expval{\ex^{-\pqty{\gWex + \gQhk - \Delta \Fnss}}}_C
  ~,
\end{align*}
where $\expval{\,\cdot\,}_C$ is the average over the trajectory class $C$.
Thus, we have Eq.~\eqref{eq:NESS_TCFT}---a TCFT for NESS systems, whose forward
and reverse processes may start in arbitrary distributions:
\begin{align}
  \frac{
    \mathcal{R}_{\mud_\textnormal{R}}\pqty{C_\textnormal{R}}
  }{
    \mathcal{P}_{\mud_\textnormal{F}}\pqty{C}
  }
  = \expval{\ex^{-\pqty{\gWex + \gQhk - \Delta \Fnss}}}_C
  \textnormal{.}
\end{align}

Now, it remains to investigate the edge cases. Suppose that either (i)
$\mathcal{P}_{\mud_\textnormal{F}}\pqty{C} = 0$ or (ii)
$\mathcal{R}_{\mud_\textnormal{R}}\pqty{C_\textnormal{R}} = 0$, but not both.
(The latter would amount to analyzing fluctuations for a \emph{pair} of
trajectories that never occur.) Since our probabilities are strictly
nonnegative, the possible behaviors of the lefthand side are either $+\infty$
or $0$, respectively, by considering the limit of a vanishing probability. In
case (i), by definition either $\ex^{-\gQhk} \to +\infty$ or $\ex^{\Delta
\Fnss} \to +\infty$ (or both) for each forward trajectory, yielding agreement
with the lefthand side. In case (ii), similarly either $\ex^{-\gQhk} \to 0$ or
$\ex^{\Delta \Fnss} \to 0$ (or both) for each forward trajectory. Since the
preceding derivation established the TCFT for all nondiverging cases, this
establishes its validity even in the divergent limiting cases.


\end{document}